\documentclass[12pt,a4paper,fleqn]{article}
\usepackage{amsmath,amssymb,amsfonts,amsthm,amsopn,graphicx}
\usepackage{tikz} 
\usetikzlibrary{fadings}
\usetikzlibrary{arrows}
\usepackage{hyperref}
\usepackage{mathrsfs} 

\usepackage[T1,T2A]{fontenc}
\usepackage[utf8]{inputenc}

\allowdisplaybreaks[3]
\newtheorem{theorem}{Теорема}[section]
\newtheorem{definition}[theorem]{Definition}

\newtheorem{proposition}[theorem]{Statement}

\numberwithin{equation}{section} 

\newcommand{\ds}{\displaystyle}

\newcommand{\bb}{{b}}

\newcommand{\opr}{\stackrel{def}{=}}

\newcommand{\Aop}{\boldsymbol{a}}
\newcommand{\aop}{\boldsymbol{b}}
\newcommand{\Kop}{\boldsymbol{k}}

\newcommand{\Alg}{\mathscr{A}}

\renewcommand{\aa}{\mathsf{a}}
\renewcommand{\bb}{\mathsf{b}}

\newcommand{\hf}{{^1\!\!/\!_2}}

\newcommand{\arrowIn}{\tikz\draw[-stealth,thick] (-1pt,0) -- (1pt,0);}

\def\GR{green!60!teal}
\def\BL{blue!90!}
\def\RD{red!90!}
\def\PU{purple!90!}
\def\MA{orange!80!}

\usepackage[body={15cm,22cm}]{geometry}

\begin{document}

\title{On integrability of a $q$-oscillator lattice with a $B$-boundary.}
\author{Sergey Sergeev
\thanks{
   Faculty of Science and Technology, 
   University of Canberra, Bruce ACT 2617, Australia
}
\thanks{ 
Department of Fundamental and Theoretical Physics,
         Research School of Physics and Engineering,
    Australian National University, Canberra, ACT 0200, Australia}
}

\date{}

\maketitle

\begin{abstract}
In this paper we propose a method of construction of a double layer-to-layer auxiliary transfer matrix defined on a half-plane with a boundary. The transfer matrix obtained has the following features:

- It produces a complete set of integrals of motion,

- Its ingredients can be seen as an auxiliary problem for $3d$ Kuniba-Okado reflection matrix,

- The model obtained has no quantum group interpretation.
\end{abstract}

\tableofcontents


\section{Introduction}

Integrable spin chains with a boundary, usually associated with Cherednik's reflection equation, have a long history, see e.g \cite{Cher:84,Skl:88}, etc.

Cherednik's reflection equation has a multidimensional generalisation -- a tetrahedral reflection equation. It has been formulated by Kulish and Isaev in \cite{IK:97}. A number of solutions to the tetrahedral reflection equation has been obtained by Kuniba and Okado  in \cite{KO:2012} (see also \cite{KOY:2019,IKT:2024} for the further development).

The aim of this paper is a step back from Kuniba and Okado results. We will focus on an auxiliary linear problem for the tetrahedral reflection equation. The auxiliary problem can be used for a formulation of a quantum integrable system on a two-dimensional layer with a boundary. We do not obtain spectral equations (it will be a subject of a separate study), however we fix a structure of an algebra of observables and some combinatorics of the layer.  As well, the auxiliary layer-to-layer transfer matrix approach allows one to establish the complete integrability of the system.

The integrable system presented here has the following features:
\begin{itemize}
\item 
Its algebra of observables is the local $q$-oscillator algebra. I consists on 
"short" $q$-oscillators $\Alg_q$ in the bulk and on "long" $q$-oscillators $\Alg_{q^2}$ on the boundary. 

From the algebraical point of view developed by Kuniba and Okado, such boundary corresponds to $B$-case.

\item We consider the simplest possible boundary choice. The system proposed seems to be an example of perhaps many relative systems.

\item Our choice of the boundary mixes rank-size principal directions of the oscillator lattice, so that the system presented has no obvious quantum group interpretation.
\end{itemize}

The paper is organised as follows. There are only two sections. We give the formulation of the model with the boundary and the proof of its complete integrability in Section  \ref{Sec2}. Then, in Section \ref{Sec3} we discuss a relation of the model with the boundary and a model on a torus and discuss a relation of our approach to the tetrahedral reflection equations.

\section{Formulation of the model}\label{Sec2}

We commence with the main definitions and with convenient graphical notations.

\subsection{$q$-oscillator algebra, $L$-operators and their graphical representation.}

The algebra of observables discussed in this paper is the local $q$-oscillator algebra.
\begin{definition}\label{q-osc}
Algebra $\Alg_{q}$ generated by $\{1,\Kop,\Kop',\Aop^{+},\Aop^{-}\}$ with the deformation parameter $q$ is defined by the following exchange relations:
\begin{equation}\label{qosc}
\Kop\Kop'=\Kop'\Kop\;,\quad
\Kop^{\#}\Aop^\pm=q^{\pm 1}\Aop^\pm \Kop^{\#},\quad
\Aop^+\Aop^-=1+q^{-1}\Kop\Kop',\quad \Aop^-\Aop^+=1+q^{}\Kop\Kop'\;.
\end{equation}
\end{definition}
If $\Kop$ and $\Kop'$ are invertible,  their ratio $\Kop'/\Kop$ becomes a center. It is often assumed, $\Kop'/\Kop=-1$, 
but this is not essential for us here since representations of the $q$-oscillator algebra will be not considered.

In what follows, three types of the $q$-oscillator algebra will appear. In addition to the standard $\Alg_q$ we will see also $\Alg_{q^2}$ and $\Alg_{-q}$, for which parameter $q$ in the definition (\ref{qosc}) is replaced to $q^2$ and to $-q$ respectively. 

Next, we define the $\Alg$-valued matrix $L[\Alg]$:
\begin{equation}\label{Lop}
L[\Alg]=
\left(\begin{array}{cccc}
L_{0,0}^{\;\;0,0} & & & \\
& L_{1,0}^{\;\;1,0} & L_{1,0}^{\;\;0,1} & \\
& L_{0,1}^{\;\;1,0} & L_{0,1}^{\;\;0,1} & \\
& & & L_{1,1}^{\;\;1,1}
\end{array}
\right)\;=\;
\left(\begin{array}{cccc} 1 &  &  &  \\
& \Kop & \Aop^+ &  \\
& \Aop^- & \Kop' &  \\
&  &  & 1\end{array}\right)\;.
\end{equation}

Last two relations of (\ref{qosc}) makes the $L$-operator rather "free fermion type $L$-matrix" than "6-vertex type $L$-matrix".

Matrix (\ref{Lop}) allows two graphical representations.  The first one is the classical Baxter's one:
\begin{equation}
\begin{tikzpicture}
\draw [-stealth, thick] (-1,0) -- (1,0);
\draw [-stealth, thick] (0,-1) -- (0,1);
\node [above] at (0,1) {\scriptsize $i'$};
\node [below] at (0,-1) {\scriptsize $i$};
\node [right] at (1,0) {\scriptsize $j'$};
\node [left] at (-1,0) {\scriptsize $j$};
\node [above] at (0.25,0) {\scriptsize $\Alg$};
\draw [fill] (0,0) circle [radius=0.07];
\node [left] at (-1.5,0) {$L_{i,j}^{\;\;i',j'}[\Alg]\;=$};
\end{tikzpicture}
\end{equation}
Sometimes it is convenient "to expand" this notation, taking the definition (\ref{Lop}) into account directly,
\begin{equation}\label{Lop2}
\begin{array}{llll}
\begin{tikzpicture}[scale=0.75]
\draw [-stealth, thin, blue] (0,-1) -- (0,1);
\draw [-stealth, thin, blue] (-1,0) -- (1,0);
\draw [fill] (0,0) circle [radius=0.07];
\node [right] at (1.25,0) {$=1$};
\end{tikzpicture} &&&\\
& 
\begin{tikzpicture}[scale=0.75]
\draw [-stealth, ultra thick] (0,-1) -- (0,1);
\draw [-stealth, thin, blue] (-1,0) -- (1,0);
\draw [fill] (0,0) circle [radius=0.07];
\node [right] at (1.25,0) {$=\Kop$};
\end{tikzpicture}
&
\begin{tikzpicture}[scale=0.75]
\draw [ultra thick] (0,-1) -- (0,0); \draw [-stealth, ultra thick] (0,0) -- (1,0);
\draw [thin, blue] (-1,0) -- (0,0); \draw [-stealth, thin, blue] (0,0) -- (0,1);
\draw [fill] (0,0) circle [radius=0.07];
\node [right] at (1.25,0) {$=\Aop^{+}$};
\end{tikzpicture}
&\\
& 
\begin{tikzpicture}[scale=0.75]
\draw [ultra thick] (-1,0) -- (0,0); \draw [-stealth, ultra thick] (0,0) -- (0,1);
\draw [thin, blue] (0,-1) -- (0,0); \draw [-stealth, thin, blue] (0,0) -- (1,0);
\draw [fill] (0,0) circle [radius=0.07];
\node [right] at (1.25,0) {$=\Aop^{-}$};
\end{tikzpicture}
&
\begin{tikzpicture}[scale=0.75]
\draw [ultra thick] (-1,0) -- (0,0); \draw [-stealth, ultra thick] (0,0) -- (1,0);
\draw [thin, blue] (0,-1) -- (0,0); \draw [-stealth, thin, blue] (0,0) -- (0,1);
\draw [fill] (0,0) circle [radius=0.07];
\node [right] at (1.25,0) {$=\Kop'$};
\end{tikzpicture}
&\\
&&&
\begin{tikzpicture}[scale=0.75]
\draw [-stealth, ultra thick] (0,-1) -- (0,1);
\draw [-stealth, ultra thick] (-1,0) -- (1,0);
\draw [fill] (0,0) circle [radius=0.07];
\node [right] at (1.25,0) {$=1$};
\end{tikzpicture}
\end{array}
\end{equation}
A square lattice in this representation corresponds to a layer-to-layer transfer matrix. Note that the $q$-oscillators inhabiting the vertices of such lattice all are independent, so that the algebra of observables is a proper tensor power of the local $q$-oscillator algebra.

Baxter's graphical representation (\ref{Lop2}) is convenient when one studies e.g. a combinatorics of the quantum lattice. However, $L$-operator (\ref{Lop}) is essentially a three dimensional object, so that the three dimensional graphical representation of $L$-operator is also required:
\begin{equation}\label{Lop3}
\begin{tikzpicture}
\draw [-stealth, ultra thick, \RD] (0,-1) -- (0,1);
\draw [-stealth, thick, \GR] (-1.4,0) -- (1.4,0);
\draw [-stealth, thick, \BL] (-1.2,-0.35) -- (1.2,0.35);
\node [left] at (-1.2,-0.4) {\scriptsize $i$};
\node [below] at (0,-1) {\scriptsize $\Alg_q$};
\node [left] at (-1.4,0) {\scriptsize $j$};
\node [above] at (1.2,0.35) {\scriptsize $i'$};
\node [right] at (1.4,0) {\scriptsize $j'$};
\node [right] at (2,0) {$=$};
\end{tikzpicture}
\begin{tikzpicture}
\draw [-stealth, ultra thick, \RD] (0,-1) -- (0,1);
\draw [-stealth, thick, \BL] (-1.4,0) -- (1.4,0);
\draw [-stealth, thick, \GR] (-1.2,-0.35) -- (1.2,0.35);
\node [left] at (-1.2,-0.4) {\scriptsize $j$};
\node [below] at (0,-1) {\scriptsize $\Alg_q$};
\node [left] at (-1.4,0) {\scriptsize $i$};
\node [above] at (1.2,0.35) {\scriptsize $j'$};
\node [right] at (1.4,0) {\scriptsize $i'$};
\node [right] at (2,0) {$=\; L_{i,j}^{\;\;i',j'}[\Alg_q]$};
\end{tikzpicture}
\end{equation}
Representation (\ref{Lop3}) is to be explained. The "first" two-dimensional space of the $L$-operator will be denoted by the blue color, and its index will be $i$ with some sub-index. The "second" space will be green and denoted by the index $j$. The red line corresponds to a representation space of the $q$-oscillator. Arrow on the lines will show the matrix or operator ordering of a product of $L$-operators. For instance, the indices $i,j$ on picture (\ref{Lop}) are incoming indices, while the indices $i',j'$ are the outgoing indices when one writes an $L$-operator formula from left to right (like in $\ds C_i^{\;\;i''}=\sum_{i'}A_{i}^{\;\;i'} B_{i'}^{\;\;i''}$).

\subsection{Monodromy matrices}

We consider a model with two layer-to-layer monodromy matrices. The best way to define them is to use the graphical representation.

The first monodromy matrix on the triangular half-plane is defined by
\begin{equation}\label{T1}
\begin{tikzpicture}[scale=0.75]
\draw [-stealth, ultra thick, \RD] (0,-1) -- (0,1);
\draw [-stealth, ultra thick, \RD] (4,-1) -- (4,1);
\draw [-stealth, ultra thick, \RD] (8,-1) -- (8,1);
\draw [-stealth, ultra thick, \RD] (12,-1) -- (12,1);
	\draw [-stealth, ultra thick, \RD] (3,0) -- (3,2);
	\draw [-stealth, ultra thick, \RD] (7,0) -- (7,2);
	\draw [-stealth, ultra thick, \RD] (11,0) -- (11,2);
		\draw [-stealth, ultra thick, \RD] (6,1) -- (6,3);
		\draw [-stealth, ultra thick, \RD] (10,1) -- (10,3);
			\draw [-stealth, ultra thick, \RD] (9,2) -- (9,4);
\draw [-stealth, thick, \BL] (-1.2,-0.4) -- (10.2,3.4); 
	\node [below] at (-1.2,-0.4) {\scriptsize $i_0$}; \node [right] at (10.2,3.4) {\scriptsize $i_0'$};
\draw [-stealth, thick, \BL] (2.8,-0.4) -- (11.2,2.4); 
	\node [below] at (2.8,-0.4) {\scriptsize $i_1$}; \node [right] at (11.2,2.4) {\scriptsize $i_1'$};
\draw [-stealth, thick, \BL] (6.8,-0.4) -- (12.2,1.4); 
	\node [below] at (6.8,-0.4) {\scriptsize $i_2$}; \node [right] at (12.2,1.4) {\scriptsize $i_2'$};
\draw [-stealth, thick, \BL] (10.8,-0.4) -- (13.2,0.4); 
	\node [below] at (10.8,-0.4) {\scriptsize $i_3$}; \node [right] at (13.2,0.4) {\scriptsize $i_3'$};
\draw [-stealth, thick, \GR] (-0.5,0.5) -- (0.5,-0.5);
	\node [above] at (-0.5,0.5) {\scriptsize $j_1$}; \node [below] at (0.7,-0.5) {\scriptsize $j_1'$};
\draw [-stealth, thick, \GR] (2.5,1.5) -- (4.5,-0.5);
	\node [above] at (2.5,1.5) {\scriptsize $j_2$}; \node [below] at (4.7,-0.5) {\scriptsize $j_2'$};
\draw [-stealth, thick, \GR] (5.5,2.5) -- (8.5,-0.5);
	\node [above] at (5.5,2.5) {\scriptsize $j_3$}; \node [below] at (8.7,-0.5) {\scriptsize $j_3'$};
\draw [-stealth, thick, \GR] (8.5,3.5) -- (12.5,-0.5);
	\node [above] at (8.5,3.5) {\scriptsize $j_4$}; \node [below] at (12.7,-0.5) {\scriptsize $j_4'$};
\node [left] at (-1,2) {$U_{\;\boldsymbol{i},\;\boldsymbol{j}}^{\;\;\boldsymbol{i}',\boldsymbol{j}'}\;=$};
\end{tikzpicture}
\end{equation}
Here in the left hand side we use vector indines
$\boldsymbol{i}\;=\;\{i_0,i_1,i_2,\dots,i_N\}$, $\boldsymbol{j}\;=\;\{j_0,j_1,j_2,\dots,j_N\}$, and so on. On this picture we use $N=4$ for simplicity, an extension to any $N$ is evident. 
All $q$-oscillators here are $\Alg_q$, in total there are $N(N+1)/2$ copies of $\Alg_q$ in the tensor product.
One could mention separately the indices $j_0,j_0'$, it could be a disconnected extra short green line to the left of $j_1,j_1'$, the delta-symbol $\delta_{j_0,j_0'}$ is implied. As well, an extra short blue line on the right is also implied, it could represent $\delta_{i_N,i_N'}$.

The second monodromy matrix is smilar, 
\begin{equation}\label{T2}
\begin{tikzpicture}[scale=0.75]
\draw [-stealth, ultra thick, \RD] (0,-1) -- (0,1);
\draw [-stealth, ultra thick, \RD] (4,-1) -- (4,1);
\draw [-stealth, ultra thick, \RD] (8,-1) -- (8,1);
\draw [-stealth, ultra thick, \RD] (12,-1) -- (12,1);
	\draw [-stealth, ultra thick, \RD] (3,0) -- (3,2);
	\draw [-stealth, ultra thick, \RD] (7,0) -- (7,2);
	\draw [-stealth, ultra thick, \RD] (11,0) -- (11,2);
		\draw [-stealth, ultra thick, \RD] (6,1) -- (6,3);
		\draw [-stealth, ultra thick, \RD] (10,1) -- (10,3);
			\draw [-stealth, ultra thick, \RD] (9,2) -- (9,4);
\draw [-stealth, thick, \GR] (-1.2,-0.4) -- (10.2,3.4); 
	\node [below] at (-1.2,-0.4) {\scriptsize ${j}_0$}; \node [right] at (10.2,3.4) {\scriptsize ${j}_0'$};
\draw [-stealth, thick, \GR] (2.8,-0.4) -- (11.2,2.4); 
	\node [below] at (2.8,-0.4) {\scriptsize ${j}_1$}; \node [right] at (11.2,2.4) {\scriptsize ${j}_1'$};
\draw [-stealth, thick, \GR] (6.8,-0.4) -- (12.2,1.4); 
	\node [below] at (6.8,-0.4) {\scriptsize ${j}_2$}; \node [right] at (12.2,1.4) {\scriptsize ${j}_2'$};
\draw [-stealth, thick, \GR] (10.8,-0.4) -- (13.2,0.4); 
	\node [below] at (10.8,-0.4) {\scriptsize ${j}_3$}; \node [right] at (13.2,0.4) {\scriptsize ${j}_3'$};
\draw [-stealth, thick, \BL] (-0.5,0.5) -- (0.5,-0.5);
	\node [above] at (-0.5,0.5) {\scriptsize ${i}_1$}; \node [below] at (0.7,-0.5) {\scriptsize ${i}_1'$};
\draw [-stealth, thick, \BL] (2.5,1.5) -- (4.5,-0.5);
	\node [above] at (2.5,1.5) {\scriptsize ${i}_2$}; \node [below] at (4.7,-0.5) {\scriptsize ${i}_2'$};
\draw [-stealth, thick, \BL] (5.5,2.5) -- (8.5,-0.5);
	\node [above] at (5.5,2.5) {\scriptsize ${i}_3$}; \node [below] at (8.7,-0.5) {\scriptsize ${i}_3'$};
\draw [-stealth, thick, \BL] (8.5,3.5) -- (12.5,-0.5);
	\node [above] at (8.5,3.5) {\scriptsize ${i}_4$}; \node [below] at (12.7,-0.5) {\scriptsize ${i}_4'$};
\node [left] at (-1,2) {$V_{\boldsymbol{i},\boldsymbol{j}}^{\;\;\boldsymbol{i}',
\boldsymbol{j}'}\;=$};
\end{tikzpicture}
\end{equation}
The difference between (\ref{T2}) and (\ref{T1}) is that blue and green lines are exchanged. In fact, it simply means the replacement $\Aop^{+}\leftrightarrow \Aop^{-}$ and $\Kop\leftrightarrow\Kop'$ (formally, this is the automorphism $\Alg_{q}\to\Alg_{q^{-1}}$).

Next, we define a boundary matrix as a tensor product of disconnected blocks:
\begin{equation}
\begin{tikzpicture}[scale=0.8]
\draw [-stealth, \PU, ultra thick] (0,-1) -- (0,1);
\node [below] at (0,-1) {\scriptsize $\Alg_{q^2}^{(k)}$};
\draw [thick, \GR] (-1,1) -- (0,0) node [sloped, pos=0.35, allow upside down]{\arrowIn}; 
\node [above] at (-1,1) {\scriptsize $j_k'''$};
\draw [thick, \GR] (0,0) -- (1,-1) node [sloped, pos=0.65, allow upside down]{\arrowIn}; 
\node [below] at (1,-1) {\scriptsize $j_{k}$};
\draw [thick, \BL] (-1,-1) -- (0,0) node [sloped, pos=0.35, allow upside down]{\arrowIn}; 
\node [below] at (-1,-1) {\scriptsize $i_{k}'$};
\draw [thick, \BL] (0,0) -- (1,1) node [sloped, pos=0.65, allow upside down]{\arrowIn}; 
\node [above] at (1,1) {\scriptsize $i_{k}''$};
\node [left] at (-2,0) {$\ds \mathop{\textrm{\Large $\otimes$}}_{k=0}^{N}$};
\node [left] at (-4,0) {$B_{\boldsymbol{i}',\boldsymbol{j}'''}^{\boldsymbol{i}'',\boldsymbol{j}}\;=$};
\end{tikzpicture}
\end{equation}
The quantum spaces in the boundary matrix all are  $\Alg_{q^2}$, this is the purple colour graphically.

Now one can define a more complicated object as follows. One can place matrix $U$ atop of matrix $V$ and glue them along the front edge with the help of matrix $B$.

Here is an elementary local block of such gluing (formally, the index notations correspond to $N=1$, and the "hidden" delta-symbols are written as the identities, e.g. $j_0''=j_0'''$, etc.):
\begin{equation}\label{skl}
\begin{tikzpicture}
\draw [-stealth, ultra thick, \RD] (0,-1) -- (0,3);
\draw [-stealth, ultra thick, \PU] (3,-1) -- (3,1);
\draw [-stealth, ultra thick, \PU] (-3,-1) -- (-3,1);
\draw [-stealth, thick, \GR] (-0.7,2.2) -- (0,2) .. controls (1.5,1.5) .. (3,0) -- (3.7,-0.6);
\node [left, \GR] at (-0.7,2.3) {\scriptsize $j_1''$};
\node [right, \GR] at (3.7,-0.7) {\scriptsize $j_1=j_1'$};
\draw [-stealth, thick, \BL] (-0.7,0.2) -- (0,0) .. controls (1.5,-0.5) .. (3,0) -- (3.7,0.4);
\node [left, \BL] at (-0.7,0.2) {\scriptsize $i_1$};
\node [right, \BL] at (3.7,0.4) {\scriptsize $i_1''=i_1'''$};
\draw [-stealth, thick, \BL] (-3.7,-0.6) -- (-3,0) .. controls (-1.5,1.5) .. (0,2) -- (0.7,2.2);
\node [left, \BL] at (-3.7,-0.7) {\scriptsize $i_0=i_0'$};
\node [right, \BL] at (0.7,2.3) {\scriptsize $i_0'''$};
\draw [-stealth, thick, \GR] (-3.7,0.4) -- (-3,0) .. controls (-1.5,-0.5) .. (0,0) -- (0.7,0.2);
\node [left, \GR] at (-3.7,0.4) {\scriptsize $j_0''=j_0'''$};
\node [right, \GR] at (0.7,0.2) {\scriptsize $j_0'$};
\node [above, \BL] at (-1.5,1.4) {\scriptsize $i_0''$};
\node [below, \BL] at (1.5,-0.4) {\scriptsize $i_1'$};
\node [above, \GR] at (1.5,1.4) {\scriptsize $j_1'''$};
\node [below, \GR] at (-1.5,-0.4) {\scriptsize $j_0$};
\node [below] at (0,-1) {\scriptsize $\Alg_{q}$};
\node [below] at (3,-1) {\scriptsize $\Alg_{q^2}^{(1)}$};
\node [below] at (-3,-1) {\scriptsize $\Alg_{q^2}^{(0)}$};
\end{tikzpicture}
\end{equation}
The same building block in the operator form is 
\begin{equation}\label{skl1}
\boldsymbol{T}_{\boldsymbol{i},\;\;\;\boldsymbol{j}''}^{\boldsymbol{i}''',\boldsymbol{j}'}\;=\;
\sum_{\boldsymbol{i}',\boldsymbol{i''},\boldsymbol{j},\boldsymbol{j}'''}
\underbrace{L_{i_1,j_0}^{i_1',j_0'}[\Alg_q]}_{V_{\boldsymbol{i},\boldsymbol{j}}^{\boldsymbol{i}',\boldsymbol{j}'}}
\underbrace{L_{i_0'',j_1''}^{i_0''',j_1'''}[\Alg_q]}_{U_{\boldsymbol{i}'',\boldsymbol{j}''}^{\boldsymbol{i}''',\boldsymbol{j}'''}}
\underbrace{L_{i_0,j_0''}^{i_0'',j_0}[\Alg_{q^2}^{(0)}] L_{i_1',j_1'''}^{i_1''',j_1'}[\Alg_{q^2}^{(1)}]}_{B_{\boldsymbol{i}',\boldsymbol{j}'''}^{\boldsymbol{i}'',\boldsymbol{j}}}
\end{equation}
Here $L_{i_0'',j_1''}^{i_0''',j_1'''}[\Alg_q]$ corresponds to a vertex from matrix $U$, formula (\ref{T1}). 
Matrix $L_{i_1,j_0}^{i_1',j_0'}[\Alg_q]$ corresponds to a vertex from matrix $V$, formula (\ref{T2}). The boundary matrices with algebras $\Alg_{q^2}$ are in purple.

The resulting picture can also be described as follows. Purple lines $\Alg_{q^2}$ are drawn on a window glass. Red lines $\Alg_q$ form a forest outside. Green and blue lines come from the forest to the window and then reflect back to the forest.

As the final step in this subsection, we define simple diagonal matrices
\begin{equation}\label{Dmat}
E(\lambda)\;=\;\left(\begin{array}{cc} 1 & 0 \\ 0 & \lambda \end{array}\right)\;,\quad
E(\lambda)_{\boldsymbol{i}}^{\boldsymbol{i}'}\;=\;\prod_{k=0}^N E(\lambda)_{i_k}^{i_k'}\;.
\end{equation}

\subsection{Transfer-matrix}

The double layers monodromy matrix $\boldsymbol{T}$ and the transfer-matrix $T(\lambda,\mu)$ are defined as follows:
\begin{equation}\label{Tmat}
\begin{array}{l}
\ds 
\boldsymbol{T}_{\boldsymbol{i},\;\;\;\boldsymbol{j}''}^{\boldsymbol{i}''',\boldsymbol{j}'}\;=\;
\sum_{\boldsymbol{i}',\boldsymbol{i''},\boldsymbol{j},\boldsymbol{j}'''}
V_{\boldsymbol{i},\boldsymbol{j}}^{\boldsymbol{i}',\boldsymbol{j}'}\,
U_{\boldsymbol{i}'',\boldsymbol{j}''}^{\boldsymbol{i}''',\boldsymbol{j}'''}\,
B_{\boldsymbol{i}',\boldsymbol{j}'''}^{\boldsymbol{i}'',\boldsymbol{j}}\\
\\
\ds 
T(\lambda,\mu) \;=\; \sum_{\boldsymbol{i},\boldsymbol{j}'',\boldsymbol{i}''',\boldsymbol{j}'}\;
\boldsymbol{T}_{\boldsymbol{i},\;\;\;\boldsymbol{j}''}^{\boldsymbol{i}''',\boldsymbol{j}'}\,
E(\lambda)_{\boldsymbol{i}'''}^{\boldsymbol{i}} E(\mu)_{\boldsymbol{j}'}^{\boldsymbol{j}''}
\end{array}
\end{equation}
Here $V$ is the lower half-plane, $U$ is the upper half-plane, $B$ is the front boundary, and $E$ are the boundary matrices in the depth. The simplest $N=1$ monodromy matrix is given by picture (\ref{skl}) and by formula (\ref{skl1}). 

Here one could comment the quasi-periodical boundary conditions in the depth. The blue lines of the upper matrix $U$, they go to the depth in the "north-east direction", then they meet matrices $E(\lambda)$, and then they come back from the "north-west" direction in the lower matrix $V$. Green lines have the similar behaviour.

Now we can formulate two main statements.
\begin{proposition}\label{prop1}
The same as (\ref{Tmat}) transfer-matrix can be written with the exchanged order of $U$ and $V$,
\begin{equation}\label{Tmat1}
\begin{array}{l}
\ds
\widetilde{\boldsymbol{T}}_{\boldsymbol{i}'',\boldsymbol{j}}^{\boldsymbol{i}',\;\boldsymbol{j}'''}\;=\;
\sum_{\boldsymbol{i}''',\boldsymbol{j}',\boldsymbol{i},\boldsymbol{j}''}\;
U_{\boldsymbol{i},\boldsymbol{j}}^{\boldsymbol{i}',\boldsymbol{j}'}\;
V_{\boldsymbol{i}'',\boldsymbol{j}''}^{\boldsymbol{i}''',\boldsymbol{j}'''}\;
B_{\boldsymbol{i}''',\boldsymbol{j}'}^{\boldsymbol{i},\boldsymbol{j}''}\;,\\
\\
\ds 
T(\lambda,\mu) \;=\; \sum_{\boldsymbol{i}'',\boldsymbol{j},\boldsymbol{i}',\boldsymbol{j}'''}
\widetilde{\boldsymbol{T}}_{\boldsymbol{i}'',\boldsymbol{j}}^{\boldsymbol{i}',\;\boldsymbol{j}'''}
E(\lambda)_{\boldsymbol{i}'}^{\boldsymbol{i}''} E(\mu)_{\boldsymbol{j}'''}^{\boldsymbol{j}}\;.
\end{array}
\end{equation}
\end{proposition}
\begin{proposition}\label{prop2}
Transfer-matrix $T(\lambda,\mu)$ is the generating function for a complete set of integrals of motion:
\begin{equation}
\biggl[ T(\lambda,\mu), T(\lambda',\mu') \biggr]\;=\;0\;.
\end{equation}
\end{proposition}
\noindent
The proof of the statements \ref{prop1} and \ref{prop2} will be given below.

\subsection{Proof of Statements \ref{prop1} and \ref{prop2}.}

The commutativity of the layer-to-layer transfer matrices follows from an auxiliary tetrahedron equation in a similar way as the commutativity of spin-chain transfer matrices follow from a Yang-Baxter equation
\cite{Maillard:1982,BS:1984,BS:2006}.

Statements \ref{prop1} and \ref{prop2} require four auxiliary tetrahedron equations.

The first two auxiliary tetrahedron equations can be represented graphically as follows:
\begin{equation}\label{TE-aux1}
\begin{tikzpicture}
\draw [-stealth, ultra thick, \BL] (-2.5,0.7) -- (-2,1.3) .. controls (-1,2.5) .. (0,3) -- (0.6,3.3);
\node [left, \BL] at (-2.5,0.5) {\scriptsize $i_2$}; 
\node [right, \BL] at (0.6,3.5) {\scriptsize $i_2''$};
\draw [-stealth, thick, \GR] (-2.7,1.6) -- (-2,1.3) .. controls (-1,0.9) .. (0,1) -- (0.6,1.1);
\node [left, \GR] at (-2.7,1.6) {\scriptsize $j_1$};
\node [right, \GR] at (0.6,1.2) {\scriptsize $j_1''$};
\draw [-stealth, thin, \BL] (-2.15,3) -- (-1.75,2.5) .. controls (-1,1.6) .. (0,1) -- (0.7,0.66);
\node [left, \BL] at (-2.15,3.2) {\scriptsize $i_1$}; 
\node [right, \BL] at (0.7,0.5) {\scriptsize $i_1''$};
\draw [-stealth, thick, \GR] (-2.15,2.3) -- (-1.75,2.5) .. controls (-0.75,3) .. (0,3) -- (0.75,3);
\node [left, \GR] at (-2.15,2.3) {\scriptsize $j_2$};
\node [right, \GR] at (0.75,2.9) {\scriptsize $j_2''$};
\draw [-stealth, ultra thick, \RD] (0,0.3) -- (0,3.7);
\draw [-stealth, ultra thick, \MA] (-1.60,3.22) -- (-2.15,0.58);
\node [below] at (0,0.3) {\scriptsize $\Alg_{q}$};
\node [above] at (-1.60,3.22) {\scriptsize $\Alg_{-q}$};
\node [right] at (2,2) {$=$};
\end{tikzpicture}
\quad\quad
\begin{tikzpicture}
\draw [-stealth, ultra thick, \BL] (-0.6,3.3) -- (0,3) .. controls (1,2.5) .. (2,1.3) -- (2.5,0.7);
\node [left, \BL] at (-0.6,3.5) {\scriptsize $i_1$};
\node [right, \BL] at (2.5,0.5) {\scriptsize $i_1''$};
\draw [-stealth, thick, \GR] (-0.6,1.1) -- (0,1) .. controls (1,0.9) .. (2,1.3) -- (2.7,1.6);
\node [left, \GR] at (-0.6,1.2) {\scriptsize $j_2$};
\node [right, \GR] at (2.7,1.7) {\scriptsize $j_2''$};
\draw [-stealth, thin, \BL] (-0.7,0.66) -- (0,1) .. controls (1,1.6) .. (1.75,2.5) -- (2.15,3);
\node [left, \BL] at (-0.7,0.5) {\scriptsize $i_2$};
\node [right, \BL] at (2.15,3.1) {\scriptsize $i_2''$};
\draw [-stealth, thick, \GR] (-0.75,3) -- (0,3) .. controls (0.75,3) .. (1.75,2.5) -- (2.15,2.3);
\node [left, \GR] at (-0.75,3) {\scriptsize $j_1$};
\node [right, \GR] at (2.15,2.3) {\scriptsize $j_1''$};
\draw [-stealth, ultra thick, \RD] (0,0.3) -- (0,3.7);
\draw [-stealth, ultra thick, \MA] (1.60,3.22) -- (2.15,0.58);
\node [below] at (0,0.3) {\scriptsize $\Alg_{q}$};
\node [above] at (1.60,3.22) {\scriptsize $\Alg_{-q}$};
\end{tikzpicture}
\end{equation}
and
\begin{equation}\label{TE-aux2}
\begin{tikzpicture}
\draw [-stealth, ultra thick, \GR] (-2.5,0.7) -- (-2,1.3) .. controls (-1,2.5) .. (0,3) -- (0.6,3.3);
\node [left, \GR] at (-2.5,0.5) {\scriptsize $j_2$}; 
\node [right, \GR] at (0.6,3.5) {\scriptsize $j_2''$};
\draw [-stealth, thick, \BL] (-2.7,1.6) -- (-2,1.3) .. controls (-1,0.9) .. (0,1) -- (0.6,1.1);
\node [left, \BL] at (-2.7,1.6) {\scriptsize $i_1$};
\node [right, \BL] at (0.6,1.2) {\scriptsize $i_1''$};
\draw [-stealth, thin, \GR] (-2.15,3) -- (-1.75,2.5) .. controls (-1,1.6) .. (0,1) -- (0.7,0.66);
\node [left, \GR] at (-2.15,3.2) {\scriptsize $j_1$}; 
\node [right, \GR] at (0.7,0.5) {\scriptsize $j_1''$};
\draw [-stealth, thick, \BL] (-2.15,2.3) -- (-1.75,2.5) .. controls (-0.75,3) .. (0,3) -- (0.75,3);
\node [left, \BL] at (-2.15,2.3) {\scriptsize $i_2$};
\node [right, \BL] at (0.75,2.9) {\scriptsize $i_2''$};
\draw [-stealth, ultra thick, \RD] (0,0.3) -- (0,3.7);
\draw [-stealth, ultra thick, \MA] (-2.15,0.58) -- (-1.60,3.22);
\node [below] at (0,0.3) {\scriptsize $\Alg_{q}$};
\node [above] at (-1.60,3.22) {\scriptsize $\Alg_{-q}$};
\node [right] at (2,2) {$=$};
\end{tikzpicture}
\quad\quad
\begin{tikzpicture}
\draw [-stealth, ultra thick, \GR] (-0.6,3.3) -- (0,3) .. controls (1,2.5) .. (2,1.3) -- (2.5,0.7);
\node [left, \GR] at (-0.6,3.5) {\scriptsize $j_1$};
\node [right, \GR] at (2.5,0.5) {\scriptsize $j_1''$};
\draw [-stealth, thick, \BL] (-0.6,1.1) -- (0,1) .. controls (1,0.9) .. (2,1.3) -- (2.7,1.6);
\node [left, \BL] at (-0.6,1.2) {\scriptsize $i_2$};
\node [right, \BL] at (2.7,1.7) {\scriptsize $i_2''$};
\draw [-stealth, thin, \GR] (-0.7,0.66) -- (0,1) .. controls (1,1.6) .. (1.75,2.5) -- (2.15,3);
\node [left, \GR] at (-0.7,0.5) {\scriptsize $j_2$};
\node [right, \GR] at (2.15,3.1) {\scriptsize $j_2''$};
\draw [-stealth, thick, \BL] (-0.75,3) -- (0,3) .. controls (0.75,3) .. (1.75,2.5) -- (2.15,2.3);
\node [left, \BL] at (-0.75,3) {\scriptsize $i_1$};
\node [right, \BL] at (2.15,2.3) {\scriptsize $i_1''$};
\draw [-stealth, ultra thick, \RD] (0,0.3) -- (0,3.7);
\draw [-stealth, ultra thick, \MA] (2.15,0.58) -- (1.60,3.22);
\node [below] at (0,0.3) {\scriptsize $\Alg_{q}$};
\node [above] at (1.60,3.22) {\scriptsize $\Alg_{-q}$};
\end{tikzpicture}
\end{equation}
Both (\ref{TE-aux1}) and (\ref{TE-aux2}) algebraically are the same,
\begin{equation}\label{TE-aux12}
\sum_{i_1',i_2',j_1',j_2'} 
M_{i_1,j_2}^{i_1',j_2'}
M_{i_2,j_1}^{i_2',j_1'}\;
L_{i_1',j_1'}^{i_1'',j_1''}
L_{i_2',j_2'}^{i_2'',j_2''}\;=\;
\sum_{i_1',i_2',j_1',j_2'} 
L_{i_2,j_2}^{i_2',j_2'}
L_{i_1,j_1}^{i_1',j_1'}\;
M_{i_2',j_1'}^{i_2'',j_1''}
M_{i_1',j_2'}^{i_1'',j_2''}\;.
\end{equation}
Here $L=L[\Alg_q]$ is the same as (\ref{Lop}),
and this $\Alg_q$ -algebra corresponds to the bright red line on the picture. In its turn
\begin{equation}\label{Mop}
M\;=\;
\left(\begin{array}{cccc}
M_{0,0}^{0,0} & 0 & 0 & M_{0,0}^{1,1} \\
0 & M_{1,0}^{1,0} & 0 & 0 \\
0 & 0 & M_{0,1}^{0,1} & 0 \\
M_{1,1}^{0,0} & 0 & 0 & M_{1,1}^{1,1}
\end{array}\right)\;=\;
\left(\begin{array}{cccc}
\Kop & 0 & 0 & \Aop^{-} \\
0 & 1 & 0 & 0 \\
0 & 0 & 1 & 0 \\
\Aop^{+} & 0 & 0 & \Kop'
\end{array}
\right)\;,\quad \Kop^{\#},\Aop^{\pm}\in \Alg_{-q}\;,
\end{equation}
and this $\Alg_{-q}$ -algebra corresponds to the orange line on the picture.

The other two auxiliary tetrahedron equations can be represented graphically as
\begin{equation}\label{TE-aux3}
\begin{tikzpicture}
\draw [-stealth, ultra thick, \BL] (-2.5,0.7) -- (-2,1.3) .. controls (-1,2.5) .. (0,3) -- (0.6,3.3);
\node [left, \BL] at (-2.5,0.5) {\scriptsize $i_2$}; 
\node [right, \BL] at (0.6,3.5) {\scriptsize $i_2''$};
\draw [-stealth, thick, \BL] (-2.7,1.6) -- (-2,1.3) .. controls (-1,0.9) .. (0,1) -- (0.6,1.1);
\node [left, \BL] at (-2.7,1.6) {\scriptsize $i_1$};
\node [right, \BL] at (0.6,1.2) {\scriptsize $i_1''$};
\draw [-stealth, thin, \GR] (-2.15,3) -- (-1.75,2.5) .. controls (-1,1.6) .. (0,1) -- (0.7,0.66);
\node [left, \GR] at (-2.15,3.2) {\scriptsize $j_1$}; 
\node [right, \GR] at (0.7,0.5) {\scriptsize $j_1''$};
\draw [-stealth, thick, \GR] (-2.15,2.3) -- (-1.75,2.5) .. controls (-0.75,3) .. (0,3) -- (0.75,3);
\node [left, \GR] at (-2.15,2.3) {\scriptsize $j_2$};
\node [right, \GR] at (0.75,2.9) {\scriptsize $j_2''$};
\draw [-stealth, ultra thick, \RD] (0,0.3) -- (0,3.7);
\node [below] at (0,0.3) {\scriptsize $\Alg_{q}$};
\draw [-stealth, ultra thick, \MA] (-2.15,0.58) -- (-1.60,3.22);
\node [above] at (-1.60,3.22) {\scriptsize $\Alg_{-q}$};
\node [right] at (2,2) {$=$};
\end{tikzpicture}
\quad\quad
\begin{tikzpicture}
\draw [-stealth, ultra thick, \GR] (-0.6,3.3) -- (0,3) .. controls (1,2.5) .. (2,1.3) -- (2.5,0.7);
\node [left, \GR] at (-0.6,3.5) {\scriptsize $j_1$};
\node [right, \GR] at (2.5,0.5) {\scriptsize $j_1''$};
\draw [-stealth, thick, \GR] (-0.6,1.1) -- (0,1) .. controls (1,0.9) .. (2,1.3) -- (2.7,1.6);
\node [left, \GR] at (-0.6,1.2) {\scriptsize $j_2$};
\node [right, \GR] at (2.7,1.7) {\scriptsize $j_2''$};
\draw [-stealth, thin, \BL] (-0.7,0.66) -- (0,1) .. controls (1,1.6) .. (1.75,2.5) -- (2.15,3);
\node [left, \BL] at (-0.7,0.5) {\scriptsize $i_2$};
\node [right, \BL] at (2.15,3.1) {\scriptsize $i_2''$};
\draw [-stealth, thick, \BL] (-0.75,3) -- (0,3) .. controls (0.75,3) .. (1.75,2.5) -- (2.15,2.3);
\node [left, \BL] at (-0.75,3) {\scriptsize $i_1$};
\node [right, \BL] at (2.15,2.3) {\scriptsize $i_1''$};
\draw [-stealth, ultra thick, \RD] (0,0.3) -- (0,3.7);
\node [below] at (0,0.3) {\scriptsize $\Alg_{q}$};
\draw [-stealth, ultra thick, \MA] (2.15,0.58) -- (1.60,3.22);
\node [above] at (1.60,3.22) {\scriptsize $\Alg_{-q}$};
\end{tikzpicture}
\end{equation}
and
\begin{equation}\label{TE-aux4}
\begin{tikzpicture}
\draw [-stealth, ultra thick, \GR] (-2.5,0.7) -- (-2,1.3) .. controls (-1,2.5) .. (0,3) -- (0.6,3.3);
\node [left, \GR] at (-2.5,0.5) {\scriptsize $j_2$}; 
\node [right, \GR] at (0.6,3.5) {\scriptsize $j_2''$};
\draw [-stealth, thick, \GR] (-2.7,1.6) -- (-2,1.3) .. controls (-1,0.9) .. (0,1) -- (0.6,1.1);
\node [left, \GR] at (-2.7,1.6) {\scriptsize $j_1$};
\node [right, \GR] at (0.6,1.2) {\scriptsize $j_1''$};
\draw [-stealth, thin, \BL] (-2.15,3) -- (-1.75,2.5) .. controls (-1,1.6) .. (0,1) -- (0.7,0.66);
\node [left, \BL] at (-2.15,3.2) {\scriptsize $i_1$}; 
\node [right, \BL] at (0.7,0.5) {\scriptsize $i_1''$};
\draw [-stealth, thick, \BL] (-2.15,2.3) -- (-1.75,2.5) .. controls (-0.75,3) .. (0,3) -- (0.75,3);
\node [left, \BL] at (-2.15,2.3) {\scriptsize $i_2$};
\node [right, \BL] at (0.75,2.9) {\scriptsize $i_2''$};
\draw [-stealth, ultra thick, \RD] (0,0.3) -- (0,3.7);
\node [below] at (0,0.3) {\scriptsize $\Alg_{q}$};
\draw [-stealth, ultra thick, \MA] (-1.60,3.22) -- (-2.15,0.58);
\node [above] at (-1.60,3.22) {\scriptsize $\Alg_{-q}$};
\node [right] at (2,2) {$=$};
\end{tikzpicture}
\quad\quad
\begin{tikzpicture}
\draw [-stealth, ultra thick, \BL] (-0.6,3.3) -- (0,3) .. controls (1,2.5) .. (2,1.3) -- (2.5,0.7);
\node [left, \BL] at (-0.6,3.5) {\scriptsize $i_1$};
\node [right, \BL] at (2.5,0.5) {\scriptsize $i_1''$};
\draw [-stealth, thick, \BL] (-0.6,1.1) -- (0,1) .. controls (1,0.9) .. (2,1.3) -- (2.7,1.6);
\node [left, \BL] at (-0.6,1.2) {\scriptsize $i_2$};
\node [right, \BL] at (2.7,1.7) {\scriptsize $i_2''$};
\draw [-stealth, thin, \GR] (-0.7,0.66) -- (0,1) .. controls (1,1.6) .. (1.75,2.5) -- (2.15,3);
\node [left, \GR] at (-0.7,0.5) {\scriptsize $j_2$};
\node [right, \GR] at (2.15,3.1) {\scriptsize $j_2''$};
\draw [-stealth, thick, \GR] (-0.75,3) -- (0,3) .. controls (0.75,3) .. (1.75,2.5) -- (2.15,2.3);
\node [left, \GR] at (-0.75,3) {\scriptsize $j_1$};
\node [right, \GR] at (2.15,2.3) {\scriptsize $j_1''$};
\draw [-stealth, ultra thick, \RD] (0,0.3) -- (0,3.7);
\node [below] at (0,0.3) {\scriptsize $\Alg_{q}$};
\draw [-stealth, ultra thick, \MA] (1.60,3.22) -- (2.15,0.58);
\node [above] at (1.60,3.22) {\scriptsize $\Alg_{-q}$};
\end{tikzpicture}
\end{equation}
Both (\ref{TE-aux3}) and (\ref{TE-aux4}) algebraically are the same,
\begin{equation}\label{TE-aux34}
\ds \sum_{i_1',i_2',j_1',j_2'} 
N_{i_1,i_2}^{i_1',i_2'}
N_{j_1,j_2}^{j_1',j_2'}
L_{i_1',j_1'}^{i_1'',j_1''}
L_{i_2',j_2'}^{i_2'',j_2''} \;=\;
\sum_{i_1',i_2',j_1',j_2'} 
L_{i_2,j_2}^{i_2',j_2'}
L_{i_1,j_1}^{i_1',j_1'}
N_{j_1',j_2'}^{j_1'',j_2''}
N_{i_1',i_2'}^{i_1'',i_2''}\;,
\end{equation}
where $L=L[\Alg_q]$, while $N=L[\Alg_{-q}]$.

Equations (\ref{TE-aux12}) and (\ref{TE-aux34}) are $2^4\times 2^4$ matrix equations, they can be verified straightforwardly.

What is important in the graphical representations, the direction of the orange line is back-to-front in (\ref{TE-aux1},\ref{TE-aux4}) and front-to-back in (\ref{TE-aux2},\ref{TE-aux3}).

The boundary on picture (\ref{skl}) requires no additional considerations. The "orange" operator $M$ just exchanges with the boundary "purple" operator $L[\Alg_{q^2}]$,
\begin{equation}\label{TR-b}
\begin{tikzpicture}
\draw [-stealth, ultra thick, \PU] (1,-1) -- (1,1);
\draw [-stealth, ultra thick, \MA] (-1,1) -- (-1,-1);
\draw [-stealth, thick, \GR] (-1.5,-0.3) -- (-1,0) .. controls (0,0.5) .. (1,0) -- (1.5,-0.3);
\draw [-stealth, thick, \BL] (-1.5,0.3) -- (-1,0) .. controls (0,-0.5) .. (1,0) -- (1.5,0.3);
\node [right] at (2.5,0) {$=$};
\node [below] at (1,-1) {\scriptsize $\Alg_{q^2}$};
\node [above] at (-1,1) {\scriptsize $\Alg_{-q}$};
\node [left, \BL] at (-1.5,0.4) {\scriptsize $i$};
\node [left, \GR] at (-1.5,-0.4) {\scriptsize $j$}; 
\node [right, \BL] at (1.5,0.4) {\scriptsize $i''$};
\node [right, \GR] at (1.5,-0.4) {\scriptsize $j''$}; 
\end{tikzpicture}
\quad\quad
\begin{tikzpicture}
\draw [-stealth, ultra thick, \PU] (-1,-1) -- (-1,1);
\draw [-stealth, ultra thick, \MA] (1,1) -- (1,-1);
\draw [-stealth, thick, \GR] (-1.5,-0.3) -- (-1,0) .. controls (0,0.5) .. (1,0) -- (1.5,-0.3);
\draw [-stealth, thick, \BL] (-1.5,0.3) -- (-1,0) .. controls (0,-0.5) .. (1,0) -- (1.5,0.3);
\node [below] at (-1,-1) {\scriptsize $\Alg_{q^2}$};
\node [above] at (1,1) {\scriptsize $\Alg_{-q}$};
\node [left, \BL] at (-1.5,0.4) {\scriptsize $i$};
\node [left, \GR] at (-1.5,-0.4) {\scriptsize $j$}; 
\node [right, \BL] at (1.5,0.4) {\scriptsize $i''$};
\node [right, \GR] at (1.5,-0.4) {\scriptsize $j''$}; 
\end{tikzpicture}
\end{equation}
\begin{equation}
\sum_{i',j'} M_{i,j}^{i',j'} \; L_{i',j'}^{i'',j''}[\Alg_{q^2}] \;=\;
\sum_{i',j'} L_{i,j}^{i',j'}[\Alg_{q^2}]\; M_{i',j'}^{i'',j''}
\end{equation}
and this exchange is trivial, $ML=LM$ due to the block-diagonal structure of $L$ and $M$. 

It follows from (\ref{TE-aux1},\ref{TR-b}), the monodromy matrices $\boldsymbol{T}=VUB$ and $\widetilde{\boldsymbol{T}}=UVB$, entering the definitions
(\ref{Tmat}) and (\ref{Tmat1}), are similar, 
\begin{equation}\label{sim}
\sum_{\boldsymbol{i}',\boldsymbol{j}'}\;
\boldsymbol{M}_{\boldsymbol{i},\boldsymbol{j}}^{\boldsymbol{i}',\boldsymbol{j'}'}\;
\boldsymbol{T}_{\boldsymbol{i}',\boldsymbol{j}'}^{\boldsymbol{i}'',\boldsymbol{j}''}\;=\;
\sum_{\boldsymbol{i}',\boldsymbol{j}'}\;
\widetilde{\boldsymbol{T}}_{\boldsymbol{i},\boldsymbol{j}}^{\boldsymbol{i}',\boldsymbol{j}'}\;
\boldsymbol{M}_{\boldsymbol{i}',\boldsymbol{j}'}^{\boldsymbol{i}'',\boldsymbol{j}''}\;,
\end{equation}
where
\begin{equation}
\boldsymbol{M}_{\boldsymbol{i},\boldsymbol{j}}^{\boldsymbol{i}',\boldsymbol{j'}'}
\;=\;
\mathop{\textrm{Trace}}_{\Alg_{-q}} \biggl(\Kop^x M_{i_N,j_N}^{i_N',j_N'} \cdots
M_{i_1,j_1}^{i_1',j_1'} M_{i_0,j_0}^{i_0',j_0'}\biggr)\;.
\end{equation}
The term $\Kop^x$ with $x>0$ makes the trace well-defined for e.g. the Fock space representation of $\Alg_{-q}$. Also, it is easy to verify, $\boldsymbol{M}$ commutes with $E(\lambda)E(\mu)$. Thus, Statement \ref{prop1} is the consequence of the similarity relation (\ref{sim}).

To prove Statement \ref{prop2}, one has to consider two pairs of the layers. The similarity can be constructed in the analogous way, with the help of two auxiliary matrices $M=M[\Alg_{-q}]$ 
and two auxiliary matrices $N=L[\Alg_{-q}]$, subject of the auxiliary tetrahedron equations (\ref{TE-aux1},\ref{TE-aux2},\ref{TE-aux3},\ref{TE-aux4}). But what happens on the boundary?

On the boundary one should use a kind of the ``reflection'' equation,
\begin{equation}\label{Refl-1}
\begin{array}{l}
\ds
\sum_{i_{\#}',i_{\#}'',j_{\#}',j_{\#}''}
M_{i_1,j_2}^{i_1',j_2'}[\Alg_{-q}^{(2)}]
L_{i_1',i_2}^{i_1'',i_2'}[\Alg_{-q}^{(1)}]
K_{12}
L_{j_1,j_2'}^{j_1',j_2''}[\Alg_{-q}^{(1)}]
M_{i_2',j_1'}^{i_2'',j_1''}[\Alg_{-q}^{(2)}] \times\\
\\
\ds \times 
L_{i_1'',j_1''}^{i_1''',j_1'''}[\Alg_{q^2}^{(0)}]
L_{i_2'',j_2''}^{i_2''',j_2'''}[\Alg_{q^2}^{(0)}] \;=\;\\
\\
\ds =\;
\sum_{i_{\#}',i_{\#}'',j_{\#}',j_{\#}''} 
L_{i_2,j_2}^{i_2',j_2'}[\Alg_{q^2}^{(0)}]
L_{i_1,j_1}^{i_1',j_1'}[\Alg_{q^2}^{(0)}]\times\\
\\
\ds \times 
M_{i_2',j_1'}^{i_2'',j_2''}[\Alg_{-q}^{(2)}]
L_{j_1'',j_2'}^{i_1''',j_2''}[\Alg_{-q}^{(1)}]
K_{12}
L_{i_1',i_2''}^{i_1'',i_2'''}[\Alg_{-q}^{(1)}]
M_{i_1'',j_2''}^{i_1''',j_2'''}[\Alg_{-q}^{(2)}]\;,
\end{array}
\end{equation}
where $K_{12}$ is the boundary operator, related to an exchange in $\Alg_{-q}^{(1)}\otimes\Alg_{-q}^{(2)}$,
\begin{equation}
\begin{array}{l}
\ds K_{12}^{} \Aop_1^{+}\;=\;q\Aop_1^{+} \frac{\Kop_2'}{\Kop_2} K_{12}\;,\quad
K_{12}^{} \Aop_1^{-}\;=\;q^{-1}\Aop_1^{-} \frac{\Kop_2}{\Kop_2'} K_{12}\;,\\
\\
\ds [K_{12}^{},\Kop_{1}^{\#}]\;=\;[K_{12}^{},\Kop_{2}^{\#}]\;=\;[K_{12}^{},\Aop_2^{\pm}]\;=\;0\;.
\end{array}
\end{equation}
Graphically, (\ref{Refl-1}) is
\begin{equation}\label{Refl-2}
\begin{tikzpicture}[scale=0.85]
\draw [-stealth, ultra thick, \PU] (0,-3) -- (0,3);\node [below] at (0,-3) {\scriptsize $\Alg_{q^2}^{(0)}$};
\draw [-stealth, thick, \BL] (-3.5,-1.5) -- (0.5,2.5);
\node [below] at (-3.5,-1.5) {\scriptsize $i_2$}; \node [above] at (0.7,2.5) {\scriptsize $i_2'''$};
\draw [-stealth, thick, \GR] (-3.5,1.5) -- (0.5,-2.5);
\node [above] at (-3.7,1.5) {\scriptsize $j_1$}; \node [below] at (0.7,-2.5) {\scriptsize $j_1'''$};
\draw [-stealth, thick, \GR] (-4.5,-0.5) -- (-3,1) .. controls (-1.5,2.5) .. (0,2) -- (0.5,1.75);
\node [below] at (-4.5,-0.5) {\scriptsize $j_2$}; \node [below] at (0.7,1.75) {\scriptsize $j_2'''$};
\draw [-stealth, thick, \BL] (-4.5,0.5) -- (-3,-1) .. controls (-1.5,-2.5) .. (0,-2) -- (0.5,-1.75);
\node [above] at (-4.5,0.5) {\scriptsize $i_1$}; \node [above] at (0.7,-1.75) {\scriptsize $i_1'''$};
\draw [-stealth, thick, \MA] (-3.2,-0.6) -- (-3,-1) .. controls (-1.5,-3.5) .. (-3,1) -- (-3.2,1.6);
\node [above] at (-3,1.6) {\scriptsize $\Alg_{-q}^{(1)}$};
\draw [-stealth, thick, \MA] (-4.2,0.4) -- (-4,0) ..controls (-1.5,-3.6) .. (-2,0) -- (-2.15,0.7);
\node [above] at (-2.15,0.7) {\scriptsize $\Alg_{-q}^{(2)}$};
\node [below] at (-1.8,-2.7) {\scriptsize $K_{12}$};
\node [right] at (1,0) {$=$};
\end{tikzpicture}
\quad
\begin{tikzpicture}[scale=0.85]
\draw [-stealth, ultra thick, \PU] (0,-3) -- (0,3);\node [below] at (0,-3) {\scriptsize $\Alg_{q^2}^{(0)}$};
\draw [-stealth, thick, \GR] (-0.5,2.5) -- (3.5,-1.5);
\node [below] at (3.5,-1.5) {\scriptsize $j_1'''$}; \node [above] at (-0.7,2.5) {\scriptsize $j_1$};
\draw [-stealth, thick, \BL] (-0.5,-2.5) -- (3.5,1.5);
\node [above] at (3.7,1.5) {\scriptsize $i_2'''$}; \node [below] at (-0.7,-2.5) {\scriptsize $i_2$};
\draw [-stealth, thick, \BL] (-0.5,1.75) -- (0,2) .. controls (1.5,2.5) .. (3,1) -- (4.5,-0.5);
\node [below] at (4.5,-0.5) {\scriptsize $i_1'''$}; \node [below] at (-0.7,1.75) {\scriptsize $i_1$};
\draw [-stealth, thick, \GR] (-0.5,-1.75) -- (0,-2) .. controls (1.5,-2.5) .. (3,-1) -- (4.5,0.5) ;
\node [above] at (4.5,0.5) {\scriptsize $j_2'''$}; \node [above] at (-0.7,-1.75) {\scriptsize $j_2$};
\draw [-stealth, thick, \MA] (3.2,-0.6) -- (3,-1) .. controls (1.5,-3.5) .. (3,1) -- (3.2,1.6);
\node [above] at (3,1.6) {\scriptsize $\Alg_{-q}^{(1)}$};
\draw [-stealth, thick, \MA] (2.15,0.7) -- (2,0) .. controls (1.5,-3.6) .. (4,0) -- (4.2,0.4);
\node [above] at (2.15,0.7) {\scriptsize $\Alg_{-q}^{(2)}$};
\node [below] at (1.8,-2.7) {\scriptsize $K_{12}$};
\end{tikzpicture}
\end{equation}
The reader can observe, the directions of the orange lines here are in accordance with the directions of the orange lines in the auxiliary tetrahedron equations.

Equation (\ref{Refl-1}) is also a $2^4\times 2^4$ matrix equation, it can be verified straightforwardly.

Combining all four auxiliary tetrahedron equations and ``reflection'' equation (\ref{Refl-1}) into a reasoning similar to that of equation (\ref{sim}), we can deduce that Statement \ref{prop2} is true.

Transfer matrix (\ref{Tmat}) has the polynomial decomposition 
\begin{equation}
T(\lambda,\mu)\;=\;\sum_{n,m=0}^{N} T_{n,m} \lambda^n \mu^m\;.
\end{equation}
Combinatorial test shows that there are $(N+1)(N+2)/2$ independent operators in the set of involutive $T_{n,m}$, what provides the completeness.

\section{Discussion}\label{Sec3}

We give in this section an alternative view to the system described in the previous section.
In particular, we establish a relation between the system and Kuniba-Okado solution of the tetrahedral reflection equation. 

\subsection{Integrable lattice on torus}

We start with a brief remainder of the standard scheme of the integrable systems in $2+1$ dimensional space-time, when a space-like surface (a layer) is a square lattice defined on a torus (simple periodical boundary conditions).

$L$-operator (\ref{Lop}) is the result of a "fermionization" of a more fundamental Korepanov's auxiliary linear problem on a two-dimensional lattice \cite{K:95},
\begin{equation}\label{Kor}
\begin{tikzpicture}
\draw [-stealth, thick] (-1,0) -- (1,0);
\node [above] at (0,1) {\scriptsize $\psi_\alpha'$};
\node [below] at (0,-1) {\scriptsize $\psi_\alpha$};
\node [right] at (1,0) {\scriptsize $\psi_\beta'$};
\node [left] at (-1,0) {\scriptsize $\psi_\beta$};
\draw [-stealth, thick] (0,-1) -- (0,1);
\draw [fill] (0,0) circle [radius=0.07];
\node [anchor=south west] at (0,0) {$\Alg_{q^2}$};
\node [right] at (2,0) {$\ds \Leftrightarrow\; \left(\psi_\alpha', \psi_\beta'\right)\;=\; 
\left( \psi_\alpha^{},\psi_\beta^{}\right)\,\cdot\,X_{\alpha,\beta}[\Alg_{q^2}]$};
\end{tikzpicture}
\end{equation}
Here $\psi_\alpha^{\#},\psi_\beta^{\#}$ are the formal auxiliary linear variables, and 
\begin{equation}
X_{\alpha,\beta}[\Alg_{q^2}]\;=\; \left(\begin{array}{cc} \Kop & \Aop^{+} \\ \Aop^{-} & \Kop'\end{array}\right)\;.
\end{equation}
The indices of matrix $X_{\alpha,\beta}$ mean that this matrix is related to the components $\alpha$ and $\beta$ in a direct sum of "one-dimensional" spaces.

We consider next an $N\times M$ square lattice
\begin{equation}
\begin{tikzpicture}
\draw [-stealth, thick] (0,-0.5) -- (0,2.5); 
\draw [-stealth, thick] (1,-0.5) -- (1,2.5);
\draw [-stealth, thick] (2,-0.5) -- (2,2.5);
	\node [below] at (0,-0.5) {\scriptsize $\psi_{\alpha_{1}}$};
	\node [below] at (1,-0.5) {\scriptsize $\psi_{\alpha_{2}}$};
	\node [below] at (2,-0.5) {\scriptsize $\psi_{\alpha_{3}}$};
	\node [above] at (0,2.5) {\scriptsize $\psi_{\alpha_{1}}'$};
	\node [above] at (1,2.5) {\scriptsize $\psi_{\alpha_{2}}'$};
	\node [above] at (2,2.5) {\scriptsize $\psi_{\alpha_{3}}'$};
\draw [-stealth, thick] (-0.5,0) -- (2.5,0);
\draw [-stealth, thick] (-0.5,1) -- (2.5,1);
\draw [-stealth, thick] (-0.5,2) -- (2.5,2);
	\node [left] at (-0.5,0) {\scriptsize $\psi_{\beta_{1}}$};
	\node [left] at (-0.5,1) {\scriptsize $\psi_{\beta_{2}}$};
	\node [left] at (-0.5,2) {\scriptsize $\psi_{\beta_{3}}$};
	\node [right] at (2.5,0) {\scriptsize $\psi_{\beta_{1}}'$};
	\node [right] at (2.5,1) {\scriptsize $\psi_{\beta_{2}}'$};
	\node [right] at (2.5,2) {\scriptsize $\psi_{\beta_{3}}'$};
\draw [fill] (0,0) circle [radius=0.05];\draw [fill] (0,1) circle [radius=0.05];\draw [fill] (0,2) circle [radius=0.05];
\draw [fill] (1,0) circle [radius=0.05];\draw [fill] (1,1) circle [radius=0.05];\draw [fill] (1,2) circle [radius=0.05];
\draw [fill] (2,0) circle [radius=0.05];\draw [fill] (2,1) circle [radius=0.05];\draw [fill] (2,2) circle [radius=0.05];
\end{tikzpicture}\qquad
\end{equation}
Here $NM$ independent copies of the algebra $\Alg_{q^2}$ inhabit the sites of the lattice. Then, the collection of relations (\ref{Kor}) allows one to write
\begin{equation}
\left(\psi_{\boldsymbol{\alpha}}',\psi_{\boldsymbol{\beta}}'\right)\;=\;
\left(\psi_{\boldsymbol{\alpha}},\psi_{\boldsymbol{\beta}}\right)\; \mathbb{X}\;,
\end{equation}
where
\begin{equation}
\psi_{\boldsymbol{\alpha}} \;=\; (\psi_{\alpha_1},\psi_{\alpha_2},\cdots,\psi_{\alpha_N})\;,\quad
\psi_{\boldsymbol{\beta}} \;=\; (\psi_{\beta_1},\psi_{\beta_2},\cdots,\psi_{\beta_M})\;,
\end{equation}
and $(N+M)\times (N+M)$ matrix $\mathbb{X}$ comprises all local matrices $X_{\alpha_i,\beta_j}[\Alg_{q^2}^{(i,j)}]$,
\begin{equation}
\mathbb{X}\;=\;X_{\alpha_1,\beta_1}[\Alg_{q^2}^{(1,1)}] 
X_{\alpha_1,\beta_2}[\Alg_{q^2}^{(1,2)}]X_{\alpha_2,\beta_1}[\Alg_{q^2}^{(2,1)}]\cdots
X_{\alpha_N,\beta_M}[\Alg_{q^2}^{(N,M)}]\;.
\end{equation}
Let next $\mathbb{I}_N$ be the identity $N\times N$ matrix, and  
\begin{equation}
E\;=\;\left(\begin{array}{cc} \lambda \mathbb{I}_N & 0 \\ 0 & \mu \mathbb{I}_M\end{array}\right)\;.
\end{equation}
Then the Korepanov determinant 
\begin{equation}\label{det}
J(\lambda,\mu)\;=\;\det\left(\mathbb{I}_{N+M} - E \mathbb{X}\right)\;=\;
\sum_{n=0}^N \sum_{m=0}^M \lambda^n \mu^m J_{n,m}
\end{equation}
is the generating function of the integrals of motion. More explicitly, formula (\ref{det}) gives the classical spectral curve $J(\lambda,\mu)=0$ of the genus $g=(N-1)(M-1)$ in the classical limit $q=1$. In the quantum $q\neq 1$ case the determinant in (\ref{det}) is to be understood as a "quantum determinant". Combinatorial details of the quantum determinant are inessential (see e.g. \cite{S:2006} and \cite{S:2009} for details), while the final prescription is simple. Each summand of the determinant corresponds to a path on the lattice, and the path components were already given by picture (\ref{Lop2}). Polynomial $J_{n,m}$ from (\ref{det}) corresponds to the sum of all closed paths on torus with the homotopy class $(n,m)$. 

The same determinant can be obtained as follows. Consider the monodromy matrix \cite{BS:2006},
\begin{equation}\label{Tmat-s0}
\begin{tikzpicture}[scale=0.75]
\draw [-stealth, thick, \GR] (-1,0) -- (10,0); 
\draw [-stealth, thick, \GR] (0,1) -- (11,1); 
\draw [-stealth, thick, \GR] (1,2) -- (12,2); 
\draw [-stealth, thick, \BL] (-0.5,-0.5) -- (2.5,2.5);
\draw [-stealth, thick, \BL] (2.5,-0.5) -- (5.5,2.5);
\draw [-stealth, thick, \BL] (5.5,-0.5) -- (8.5,2.5);
\draw [-stealth, thick, \BL] (8.5,-0.5) -- (11.5,2.5);
\draw [-stealth, ultra thick, \PU] (0,-0.5) -- (0,0.5);
\draw [-stealth, ultra thick, \PU] (3,-0.5) -- (3,0.5);
\draw [-stealth, ultra thick, \PU] (6,-0.5) -- (6,0.5);
\draw [-stealth, ultra thick, \PU] (9,-0.5) -- (9,0.5);
\draw [-stealth, ultra thick, \PU] (1,0.5) -- (1,1.5);
\draw [-stealth, ultra thick, \PU] (4,0.5) -- (4,1.5);
\draw [-stealth, ultra thick, \PU] (7,0.5) -- (7,1.5);
\draw [-stealth, ultra thick, \PU] (10,0.5) -- (10,1.5);
\draw [-stealth, ultra thick, \PU] (2,1.5) -- (2,2.5);
\draw [-stealth, ultra thick, \PU] (5,1.5) -- (5,2.5);
\draw [-stealth, ultra thick, \PU] (8,1.5) -- (8,2.5);
\draw [-stealth, ultra thick, \PU] (11,1.5) -- (11,2.5);
\node [below] at (-0.7,-0.5) {\scriptsize $i_1$};\node [above] at (2.8,2.5) {\scriptsize $i_1'$};
\node [below] at (2.3,-0.5) {\scriptsize $i_2$};\node [above] at (5.8,2.5) {\scriptsize $i_2'$};
\node [below] at (5.3,-0.5) {\scriptsize $\cdots$};\node [above] at (8.8,2.5) {\scriptsize $\cdots$};
\node [below] at (8.3,-0.5) {\scriptsize $i_N$};\node [above] at (11.8,2.5) {\scriptsize $i_N'$};
\node [left] at (-1,0) {\scriptsize $j_1$}; \node [right] at (10,0) {\scriptsize $j_1'$};
\node [left] at (0,1) {\scriptsize $j_2$}; \node [right] at (11,1) {\scriptsize $j_2'$};
\node [left] at (1,2) {\scriptsize $j_M$}; \node [right] at (12,2) {\scriptsize $j_M'$};
\node [left] at (-1,1) {$T_{\boldsymbol{i},\boldsymbol{j}}^{\boldsymbol{i}',\boldsymbol{j}'}\;=$};
\end{tikzpicture}
\end{equation}
where $L[\Alg_{q^2}]$ are defined by (\ref{Lop},\ref{Lop3}), and consider its trace,
\begin{equation}\label{Tmat-s}
T(\lambda,\mu) \;=\; \sum_{\boldsymbol{i}^{\#},\boldsymbol{j}^{\#}} T_{\boldsymbol{i},\boldsymbol{j}}^{\boldsymbol{i}',\boldsymbol{j}'} E(\lambda)_{\boldsymbol{i}'}^{\boldsymbol{i}} E(\mu)_{\boldsymbol{j}'}^{\boldsymbol{j}} \;=\;
\sum_{n=0}^N \sum_{m=0}^M \lambda^n \mu^m T_{n,m}\;,
\end{equation}
where $E(\lambda),E(\mu)$ are given by (\ref{Dmat}). Then the relation between the decompositions (\ref{det}) 
and (\ref{Tmat-s}) is 
\begin{equation}
T_{n,m}\;=\;(-)^{nm+n+m} J_{n,m}\;,
\end{equation} 
where the "fermionic" sign counts the number of closed loops on the torus.

The commutativity of the transfer-matrices (\ref{Tmat-s}) follows from an auxiliary tetrahedron equation
\begin{equation}\label{TE-aux00}
\begin{array}{l}
\ds \sum_{i_1',i_2',j_1',j_2'} 
L_{i_1,i_2}^{i_1',i_2'}[\Alg^{(1)}_{-q^2}] 
L_{j_1,j_2}^{j_1',j_2'}[\Alg^{(1)}_{-q^2}] 
L_{i_1',j_1'}^{i_1'',j_1''}[\Alg^{(2)}_{q^2}] 
L_{i_2',j_2'}^{i_2'',j_2''}[\Alg^{(2)}_{q^2}]\;=\\
\\
\ds =\;
\ds \sum_{i_1',i_2',j_1',j_2'} 
L_{i_2,j_2}^{i_2',j_2'}[\Alg^{(2)}_{q^2}]
L_{i_1,j_1}^{i_1',j_1'}[\Alg^{(2)}_{q^2}]
L_{j_1',j_2'}^{j_1'',j_2''}[\Alg^{(1)}_{-q^2}]
L_{i_1',i_2'}^{i_1'',i_2''}[\Alg^{(1)}_{-q^2}]\;,
\end{array}
\end{equation}
where $\Alg_{-q^2}^{(1)}=\Alg_{-q^2}\otimes 1$ and $\Alg_{q^2}^{(2)}=1\otimes\Alg_{q^2}$, 
which can be verified straightforwardly. Graphically equation (\ref{TE-aux00}) is represented in usual way, see (\ref{TE-aux3}):
\begin{equation}\label{TE-aux10}
\begin{tikzpicture}
\draw [-stealth, ultra thick, \BL] (-2.5,0.7) -- (-2,1.3) .. controls (-1,2.5) .. (0,3) -- (0.6,3.3);
\node [left, \BL] at (-2.5,0.5) {\scriptsize $i_2$}; 
\node [right, \BL] at (0.6,3.5) {\scriptsize $i_2''$};
\draw [-stealth, thick, \BL] (-2.7,1.6) -- (-2,1.3) .. controls (-1,0.9) .. (0,1) -- (0.6,1.1);
\node [left, \BL] at (-2.7,1.6) {\scriptsize $i_1$};
\node [right, \BL] at (0.6,1.2) {\scriptsize $i_1''$};
\draw [-stealth, thin, \GR] (-2.15,3) -- (-1.75,2.5) .. controls (-1,1.6) .. (0,1) -- (0.7,0.66);
\node [left, \GR] at (-2.15,3.2) {\scriptsize $j_1$}; 
\node [right, \GR] at (0.7,0.5) {\scriptsize $j_1''$};
\draw [-stealth, thick, \GR] (-2.15,2.3) -- (-1.75,2.5) .. controls (-0.75,3) .. (0,3) -- (0.75,3);
\node [left, \GR] at (-2.15,2.3) {\scriptsize $j_2$};
\node [right, \GR] at (0.75,2.9) {\scriptsize $j_2''$};
\draw [-stealth, ultra thick, \PU] (0,0.3) -- (0,3.7);
\node [below] at (0,0.3) {\scriptsize $\Alg_{q^2}$};
\draw [-stealth, ultra thick, \MA] (-2.15,0.58) -- (-1.60,3.22);
\node [above] at (-1.60,3.22) {\scriptsize $\Alg_{-q^2}$};
\node [right] at (2,2) {$=$};
\end{tikzpicture}
\quad\quad
\begin{tikzpicture}
\draw [-stealth, ultra thick, \GR] (-0.6,3.3) -- (0,3) .. controls (1,2.5) .. (2,1.3) -- (2.5,0.7);
\node [left, \GR] at (-0.6,3.5) {\scriptsize $j_1$};
\node [right, \GR] at (2.5,0.5) {\scriptsize $j_1''$};
\draw [-stealth, thick, \GR] (-0.6,1.1) -- (0,1) .. controls (1,0.9) .. (2,1.3) -- (2.7,1.6);
\node [left, \GR] at (-0.6,1.2) {\scriptsize $j_2$};
\node [right, \GR] at (2.7,1.7) {\scriptsize $j_2''$};
\draw [-stealth, thin, \BL] (-0.7,0.66) -- (0,1) .. controls (1,1.6) .. (1.75,2.5) -- (2.15,3);
\node [left, \BL] at (-0.7,0.5) {\scriptsize $i_2$};
\node [right, \BL] at (2.15,3.1) {\scriptsize $i_2''$};
\draw [-stealth, thick, \BL] (-0.75,3) -- (0,3) .. controls (0.75,3) .. (1.75,2.5) -- (2.15,2.3);
\node [left, \BL] at (-0.75,3) {\scriptsize $i_1$};
\node [right, \BL] at (2.15,2.3) {\scriptsize $i_1''$};
\draw [-stealth, ultra thick, \PU] (0,0.3) -- (0,3.7);
\node [below] at (0,0.3) {\scriptsize $\Alg_{q^2}$};
\draw [-stealth, ultra thick, \MA] (2.15,0.58) -- (1.60,3.22);
\node [above] at (1.60,3.22) {\scriptsize $\Alg_{-q^2}$};
\end{tikzpicture}
\end{equation}
In this picture the blue-green rule formulated after picture (\ref{Lop3}) is broken for auxiliary operators $L[\Alg_{-q^2}]$, while this rule is valid for transfer-matrix (\ref{Tmat-s0}).

The quantum lattice described in this subsection corresponds to $\mathcal{U}_{q^2}(\widehat{sl}_M)$ chain of the length $N$, or to $\mathcal{U}_{q^2}(\widehat{sl}_N)$ chain of the length $M$, what is known as the rank-size duality \cite{BS:2006}. Spectral equations for such lattice are the nested Bethe Ansatz equations \cite{S:2006}.

\subsection{Korepanov's auxiliary linear problem}

Matrices $X_{\alpha,\beta}[\Alg]$ and $L_{\alpha,\beta}[\Alg]$ are used for a construction of quantum maps (i.e $3d$ $R$-matrices and other intertwining objects).

An example is the fundamental intertwining relation,  
\begin{equation}
\begin{tikzpicture}
\draw [-stealth, thick] (-1.5,-0.5) -- (0.5,1.5); \node [below] at (-1.5,-0.5) {\scriptsize $\gamma$};
\draw [-stealth, thick] (0.5,-1.5) -- (-1.5,0.5); \node [below] at (0,-2) {\scriptsize $\beta$};
\draw [-stealth, thick] (0,-2) -- (0,2); \node [below] at (0.5,-1.5) {\scriptsize $\alpha$};
\draw [fill] (0,-1) circle [radius=0.07];
\draw [fill] (0,1) circle [radius=0.07];
\draw [fill] (-1,0) circle [radius=0.07];
\node [right] at (0,1) {\scriptsize $\Alg_{q^2}^{(3)}$};
\node [right] at (0,-1) {\scriptsize $\Alg_{q^2}^{(1)}$};
\node [above] at (-1,0.1) {\scriptsize $\Alg_{q^2}^{(2)}$};
\node [left] at (-2,0) {$R_{123}$};
\node [left] at (2.3,0) {$=$};
\end{tikzpicture}\qquad\;
\begin{tikzpicture}
\draw [-stealth, thick] (1.5,-0.5) -- (-0.5,1.5); \node [below] at (1.5,-0.5) {\scriptsize $\alpha$};
\draw [-stealth, thick] (-0.5,-1.5) -- (1.5,0.5); \node [below] at (0,-2) {\scriptsize $\beta$};
\draw [-stealth, thick] (0,-2) -- (0,2); \node [below] at (-0.5,-1.5) {\scriptsize $\gamma$};
\draw [fill] (0,-1) circle [radius=0.07];
\draw [fill] (0,1) circle [radius=0.07];
\draw [fill] (1,0) circle [radius=0.07];
\node [right] at (0,1) {\scriptsize $\Alg_{q^2}^{(1)}$};
\node [right] at (0,-1) {\scriptsize $\Alg_{q^2}^{(3)}$};
\node [above] at (1,0.1) {\scriptsize $\Alg_{q^2}^{(2)}$};
\node [right] at (2,0) {$R_{123}$};
\end{tikzpicture}\qquad
\end{equation}
which algebraically is either the quantum Korepanov equation (same as Karpanov-Voevodsky equation \cite{KV:91,KS:93})
\begin{equation}\label{RXXX}
R_{123} \; 
X_{\alpha,\beta}^{(1)}
X_{\alpha,\gamma}^{(2)}
X_{\beta,\gamma}^{(3)}\;=\;
X_{\beta,\gamma}^{(3)}
X_{\alpha,\gamma}^{(2)}
X_{\alpha,\beta}^{(1)}\; R_{123}\;,
\end{equation}
or the quantum local Yang-Baxter equation \cite{BS:2006},
\begin{equation}\label{RLLL}
R_{123} \; 
L_{\alpha,\beta}^{(1)}
L_{\alpha,\gamma}^{(2)}
L_{\beta,\gamma}^{(3)}\;=\;
L_{\beta,\gamma}^{(3)}
L_{\alpha,\gamma}^{(2)}
L_{\alpha,\beta}^{(1)}\; R_{123}\;,
\end{equation}
where $X_{\alpha,\beta}^{(j)}=X_{\alpha,\beta}[\Alg_{q^2}^{(j)}]$, 
$L_{\alpha,\beta}^{(j)}=L_{\alpha,\beta}[\Alg_{q^2}^{(j)}]$, $\Alg_{q^2}^{(1)}=\Alg_{q^2}\otimes 1\otimes 1$, $\Alg_{q^2}^{(2)}=1\otimes \Alg_{q^2}\otimes 1$, $\Alg_{q^2}^{(3)}=1\otimes 1 \otimes \Alg_{q^2}$, and 
$R_{123}$ is the $3d$ $R$-matrix of the tetrahedron equation, its matrix elements are unambiguously defined by relation (\ref{RXXX}) or by equivalent relation (\ref{RLLL}). This is known for a long time.
\\

Now, let us consider the following configuraion:
\begin{equation}\label{zag-0}
\begin{array}{l}
\ds s_L\;=\; X_{\beta\gamma}^{(2)} X_{\beta\delta}^{(\bb)} X_{\alpha\gamma}^{(\bb')}
X_{\alpha\delta}^{(1)} X_{\alpha\beta}^{(\aa)}X_{\gamma\delta}^{(\aa')}\;,\\
\\
\ds s_R\;=\;X_{\alpha\beta}^{(\aa)} X_{\gamma\delta}^{(\aa')} X_{\alpha\delta}^{(1)} 
X_{\beta\delta}^{(\bb)}X_{\alpha\gamma}^{(\bb')} X_{\beta\gamma}^{(2)}\;.
\end{array}
\end{equation}
Graphically, configuration $s_L$ looks like 
\begin{equation}
\begin{tikzpicture}[scale=0.7]
\draw[black,thick,-stealth] (-2,-1) -- (6,3) node[right] {\scriptsize $\beta$};
\draw[black,thick,-stealth] (-2,1) -- (6,-3) node[right] {\scriptsize $\gamma$};
\draw[black,thick,-stealth] (1,-4) -- (5,4) node[above] {\scriptsize $\alpha$};
\draw[black,thick,-stealth] (1,4) -- (5,-4) node[below] {\scriptsize $\delta$};
\filldraw[black] (0,0) circle (2pt);
	\node[above] at (0,0) {\scriptsize $\Alg_{q^2}^{(2)}$};
\filldraw[black] (3,0) circle (2pt); 
	\node[right]  at (3,0) {\scriptsize $\Alg_{q^2}^{(1)}$};
\filldraw[black] (4,2) circle (2pt); 
	\node[right] at (3.8,1.8) {\scriptsize $\Alg_{q^2}^{(\aa)}$};
\filldraw[black] (4,-2) circle (2pt); 
	\node[above right] at (3.8,-2.2) {\scriptsize $\Alg_{q^2}^{(\aa')}$};
\filldraw[black] (2.4,1.2) circle (2pt); 
	\node[left] at (2.3,1.3) {\scriptsize $\Alg_{q^2}^{(\bb)}$};
\filldraw[black] (2.4,-1.2) circle (2pt); 
	\node[left] at (2.5,-1.5) {\scriptsize $\Alg_{q^2}^{(\bb')}$};
\node [left] at (-3,0) {$s_L\;=$};
\end{tikzpicture}
\end{equation}
Configurations $s_L$ and $s_R$ can be intertwined, 
\begin{equation}\label{KS1}
K\; s_L\;=\; s_R \; K\;,
\end{equation}
where
\begin{equation}\label{K1}
K\;=\;R_{\aa,1,\bb}^{-1} R_{\aa,\bb',2}^{-1} R_{2,\bb,\aa'}^{} R_{\bb',1,\aa'}^{}\;=\;
R_{\bb',1,\aa'}^{} R_{2,\bb,\aa'}^{} R_{\aa,\bb',2}^{-1} R_{\aa,1,\bb}^{-1}\;.
\end{equation} 
Here we do not use here the symmetries of $q$-oscillator $R$-matrix writing $R$ or $R^{-1}$ in accordance to  (\ref{RXXX}).

This paper is based in fact on certain observations made for the intertwining relation (\ref{KS1}).

\subsection{Classical limit}

Consider relation (\ref{KS1}) in the classical limit $q\to 1$. For instance, relation (\ref{RLLL}) becomes the local Yang-Baxter equation in this limit (equation (\ref{RXXX}) becomes just Korepanov equation). The quantum $q$-oscillator algebras (\ref{qosc}) are replaced by the Poisson algebras in the classical limit,
\begin{equation}\label{qpois}
\Alg_{q^n}\;:\;\;\{\Kop^{\#},\Aop^{\pm}\}\;=\;\pm n \Kop^{\#}\Aop^{\pm}\;,\quad
\{\Aop^{+},\Aop^{-}\}\;=\;n\Kop\Kop'\;,\quad \Aop^{+}\Aop^{-}+\Kop\Kop'\;=\;1\;.
\end{equation}
Equation (\ref{KS1}) in the classical limit 
\begin{equation}\label{KS2}
s_L^{}\;=\;s_R'\;,\quad s_R'\; \opr\; K^{-1}s_R^{} K\;,
\end{equation}
defines a symplectic map of six classical Poisson algebras (\ref{qpois}) $\Alg_{q^2}^{\otimes 6}\to \Alg_{q^2}^{\otimes 6}$.
Then, there is the first observation:
\begin{proposition} 
If one takes the classical map $\mathcal{K}\;:\; s_L^{} \to s_R'$ and puts here "by hands" $\Alg^{(\aa')}=\Alg^{(\aa)}$ and $\Alg^{(\bb')}=\Alg^{(\bb)}$ in $s_L$, then $\mathcal{K}$ provides $\Alg^{(\aa')\prime}=\Alg^{(\aa)\prime}$ and $\Alg^{(\bb')\prime}=\Alg^{(\bb)\prime}$ in $s_R'$.
\end{proposition} 
The related observation is 
\begin{proposition} If $\Alg^{(\aa')}=\Alg^{(\aa)}$ and $\Alg^{(\bb')}=\Alg^{\bb}$,  then the map $\mathcal{K}$, defined by (\ref{KS2}), is also symplectic, but the Poisson algebra cnahges:
\begin{equation}
\Alg^{(1)}=\Alg_{q^2}^{(1)},\;\Alg^{(2)}=\Alg_{q^2}^{(2)}\;\;\;\textrm{but}\;\;\; 
\Alg^{(\aa)}=\Alg_{q}^{(\aa)},\;\Alg^{(\bb)}=\Alg_{q}^{(\bb)}\;.
\end{equation}
\end{proposition}
The reason of this subsection is that it is easy to explain on the classical level why the $q$-oscillator algebra is changed after the identification. Consider two classical algebras, $\Alg_{q^2}^{(\aa_1)}$ and $\Alg_{q^2}^{(\aa_2)}$, with the standard bracket 
\begin{equation}\label{Pois-2}
\{\Aop^{+},\Aop^{-}\}=2(1-\Aop^{+}\Aop^{-})\;.
\end{equation}
Making then an elementary change of variables, for instance 
\begin{equation}\label{cl-ex}
\Aop^{\pm}_{\aa}\;=\;\frac{1}{2}(\Aop^{\pm}_{\aa_1}+\Aop^{\pm}_{\aa_2})\;,\quad
\aop^{\pm}_{\aa}\;=\;\frac{1}{2}(\Aop^{\pm}_{\aa_1}-\Aop^{\pm}_{\aa_2})\;\,
\end{equation}
then one obtains
\begin{equation}
\{\Aop^{+}_{\aa},\Aop^{-}_{\aa}\}\;=\;( 1- \Aop^{+}_{\aa}\Aop^{-}_{\aa} - \aop^{+}_{\aa}\aop^{-}_{\aa})\;.
\end{equation}
The identification condition $\aop_{\aa}^{\pm}=0$ gives the new bracket,
\begin{equation}\label{Pois-1}
\{\Aop^{+}_{\aa},\Aop^{-}_{\aa}\}\;=\;( 1- \Aop^{+}_{\aa}\Aop^{-}_{\aa})\;.
\end{equation}
Comparing (\ref{Pois-2}) and (\ref{Pois-1}), we see that the Poisson algebra is changed, $\Alg_{q^2}^{(\aa_1)}\otimes\Alg_{q^2}^{(\aa_2)}\to\Alg_{q}^{(\aa)}$.

\subsection{Quantum reflection map}

Consider now the configurations (\ref{zag-0}) with identified $\Alg^{(\aa')}=\Alg^{(\aa)}$ and $\Alg^{(\bb')}=\Alg^{(\bb)}$, 
\begin{equation}\label{zag-1}
\begin{array}{l}
\ds \tilde{s}_L\;=\; 
X_{\beta\gamma}^{(2)} 
X_{\beta\delta}^{(\bb)} 
X_{\alpha\gamma}^{(\bb)}
X_{\alpha\delta}^{(1)} 
X_{\alpha\beta}^{(\aa)} 
X_{\gamma\delta}^{(\aa)}\;,\\
\\
\ds \tilde{s}_R\;=\;
X_{\alpha\beta}^{(\aa)} 
X_{\gamma\delta}^{(\aa)} 
X_{\alpha\delta}^{(1)}
X_{\beta\delta}^{(\bb)}
X_{\alpha\gamma}^{(\bb)}
X_{\beta\gamma}^{(2)}\;.
\end{array}
\end{equation}
A proper graphical representation for e.g. $\tilde{s}_L$ is
\begin{equation}\label{mirror}
\begin{tikzpicture}[scale=0.7]
\fill [blue!20!, path fading = south] (-3,0) -- (7,0) -- (7,-5) -- (-3,-5) -- cycle;
\draw[black,thick,-stealth] (-2,-1) -- (6,3) node[right] {\scriptsize $\beta$};
\draw[black,thick,-stealth] (-2,1) -- (6,-3) node[right] {\scriptsize $\gamma$};
\draw[black,thick,-stealth] (1,-4) -- (5,4) node[above] {\scriptsize $\alpha$};
\draw[black,thick,-stealth] (1,4) -- (5,-4) node[below] {\scriptsize $\delta$};
\filldraw[black] (0,0) circle (2pt) node[above]  {\scriptsize $\Alg^{(2)}$};
\filldraw[black] (3,0) circle (2pt) node[right]  {\scriptsize $\Alg^{(1)}$};
\filldraw[black] (4,2) circle (2pt) node[below right] {\scriptsize $\Alg^{(\aa)}$};
\filldraw[black] (4,-2) circle (2pt) node[above right] {\scriptsize $\Alg^{(\aa)}$};
\filldraw[black] (2.4,1.2) circle (2pt) node[above left] {\scriptsize $\Alg^{(\bb)}\;$};
\filldraw[black] (2.4,-1.2) circle (2pt) node[below left] {\scriptsize $\Alg^{(\bb)}\;$};
\node [left] at (-3.3,0) {$\tilde{s}_L\;=$};
\draw [thin, blue] (-3,0) -- (7,0);
\end{tikzpicture}
\end{equation}
What stands above the horizontal line, it is our world. What stands below the horizontal line, it is the world ``through the looking-glass'''. Algebras of observables in our world and through the looking glass coincide, while the algebras on the looking glass stay separately. In our case they are
\begin{equation}\label{balg}
\Alg^{(\aa)}\;=\;\Alg_q^{(\aa)}\;,\;\;
\Alg^{(\bb)}\;=\;\Alg_q^{(\bb)}\;,\;\;
\Alg^{(1)}\;=\;\Alg_{q^2}^{(1)}\;,\;\;
\Alg^{(2)}\;=\;\Alg_{q^2}^{(2)}\;.
\end{equation}
The following statement is true:
\begin{proposition}
Quantum intetwining relation 
\begin{equation}\label{KS3}
K_{1,2;\aa,\bb} \; \tilde{s}_L \;=\; \tilde{s}_R\; K_{1,2;\aa,\bb}
\end{equation}
for the algebras (\ref{balg}) is well defined and has an unique solution.
\end{proposition}
\noindent
Equation (\ref{KS3}), being rewritten in components\footnote{The advantabe of Korepanov's matrices $X$ is that $X_{\alpha,\beta}[\Alg] X_{\gamma,\delta}[\Alg]=X_{\gamma\delta}[\Alg] X_{\alpha,\beta}[\Alg]$. It is not so for the matrices $L_{\alpha,\beta}[\Alg]$.}, coincides with the exchange relations for Kuniba-Okado reflection $K$-matrix, what proves the statement. Note also, the ``real world'' side of configuration (\ref{mirror}) literally coincides with the $K$-matrix configuration considered in \cite{IK:97}, while the algebraical structure of $\widetilde{s}_L$ and $\widetilde{s}_R$ is nothing but the quantised algebra of functions on $\boldsymbol{U}_q(B_2^{(1)})\sim\boldsymbol{U}_q(C_2^{(1)})$ \cite{KO:2012,KOY:2019,Soi:90}.

\subsection{Alternative form for configurations (\ref{zag-1}) }

The idea of the identification of the algebras on both sides of the looking glass provides "short" algebras $\Alg_{q}$ in the bulk and "long" algebras $\Alg_{q^2}$ on the boundary. It corresponds to $B$ type algebras in the algebraic approach to the reflection equation \cite{KO:2012}. However, the same relation (\ref{KS3}) can be rewritten in a $C$-form.

Namely, configurations (\ref{zag-1}) can be rewritten in the form
\begin{equation}
\tilde{s}_{L,R}\;\to\; P_{\alpha,\delta} \tilde{s}_{L,R} P_{\alpha,\delta} P_{\alpha,\beta} P_{\gamma,\delta} P_{\alpha,\gamma} P_{\beta,\delta}\;,
\end{equation}
where $P_{\alpha,\beta}$, etc. are the permutation matrices in the direct sum of the spaces $\alpha,\beta,\gamma,\delta$. For example,
\begin{equation}
P_{\alpha,\beta}\;=\;\left(\begin{array}{cccc}
0 & 1 & 0 & 0\\
1 & 0 & 0 & 0\\
0 & 0 & 1 & 0\\
0 & 0 & 0 & 1
\end{array}\right)\;,\quad
P_{\alpha,\gamma}\;=\;\left(\begin{array}{cccc}
0 & 0 & 1 & 0\\
0 & 1 & 0 & 0\\
1 & 0 & 0 & 0\\
0 & 0 & 0 & 1
\end{array}\right)\;,\quad \textrm{etc.}
\end{equation}
As the result, matrices (\ref{zag-1}) become
\begin{equation}\label{zag-2}
\begin{array}{l}
\ds \tilde{s}_L\;=\; X_{\beta,\gamma}^{(2)} \; Y^{(\bb)}  \; X_{\gamma,\beta}^{(1)}\; Y^{(\aa)}\;,\\
\\
\ds \tilde{s}_R\;=\; Y^{(\aa)} \; X_{\beta,\gamma}^{(1)} \; Y^{(\bb)} \; X_{\gamma,\beta}^{(2)} \;,
\end{array}
\end{equation}
where
\begin{equation}
\begin{array}{l}
\ds Y^{(\aa)}\;=\;X_{\delta,\beta}^{(\aa)}\; X_{\gamma,\alpha}^{(\aa)} P_{\alpha,\gamma}P_{\beta,\delta}
\;=\;
\left(\begin{array}{cccc}
\Aop^{-}_{\aa} & 0 & \Kop_{\aa}' & 0 \\
0 & \Aop^{-}_{\aa} & 0 & \Kop_{\aa}' \\
\Kop_{\aa}^{} & 0 & \Aop^{+}_{\aa} & 0 \\
0 & \Kop_{\aa}^{} & 0 & \Aop^{+}_{\aa}
\end{array}\right)\;,\\
\\
\ds Y^{(\bb)}\;=\;X_{\delta,\gamma}^{(\bb)} X_{\beta,\alpha}^{(\bb)} P_{\alpha,\beta} P_{\gamma,\delta}\;=\;
\left(\begin{array}{cccc}
\Aop^{-}_{\bb} & \Kop_{\bb}' & 0 & 0 \\
\Kop_{\bb}^{} & \Aop^{+}_{\bb} & 0 & 0 \\
0 & 0 & \Aop^{-}_{\bb} & \Kop_{\bb}' \\
0 & 0 & \Kop_{\bb}^{} & \Aop^{+}_{\bb}
\end{array}\right)\;,
\end{array}
\end{equation}
and
\begin{equation}
X_{\beta,\gamma}[\Alg]\;=\;
\left(\begin{array}{cccc}
1 & 0 & 0 & 0 \\
0 & \Kop & \Aop^+ & 0 \\
0 & \Aop^- & \Kop' & 0 \\
0 & 0 & 0 & 1
\end{array}\right)\;,\quad
X_{\gamma,\beta}[\Alg]\;=\;
\left(\begin{array}{cccc}
1 & 0 & 0 & 0 \\
0 & \Kop' & \Aop^- & 0 \\
0 & \Aop^+ & \Kop & 0 \\
0 & 0 & 0 & 1
\end{array}\right)\;.
\end{equation}
These are the matrices written in the direct sum of four one-dimensional spaces. However, the same matrices can be understood as the matrices in the direct product of two two-dimensional  spaces in the same way as in (\ref{Lop},\ref{Lop2}):
\begin{equation}
\begin{array}{l}
\ds Y^{(\aa)} \;\to\; 1\otimes Y[\Alg^{(\aa)}]\;\stackrel{def}{=}\;Y_\beta[\Alg^{(\aa)}]\;,\quad Y^{(\bb)} \;\to\; Y[\Alg^{(\bb)}]\otimes 1\;\stackrel{def}{=}\;Y_\alpha[\Alg^{(\bb)}]\;,\\
\\
\ds 
X_{\beta,\gamma}[\Alg]\;\to\;L_{\alpha,\beta}[\Alg]\;,\quad
X_{\gamma,\beta}[\Alg]\;\to\;L_{\beta,\alpha}[\Alg]\;.
\end{array}
\end{equation}
Here matrix $Y$ is introduced in addition to matrix (\ref{Lop2}), 
\begin{equation}\label{Y}
Y[\Alg]\;=\;
\left(\begin{array}{cc}
Y_0^0 & Y_0^1\\
Y_1^0 & Y_1^1
\end{array}\right)\;=\;
\left(\begin{array}{cc}
\Aop^- & \Kop' \\
\Kop & \Aop^+
\end{array}\right)\;.
\end{equation}
Matrix $Y$ has the evident graphical representation:
\begin{equation}
\begin{tikzpicture}
\draw [thick] (-1,1) -- (0,0);
\draw [-stealth, thick] (0,0) -- (1,1);
\draw [blue] (-1.5,0) -- (1.5,0);
\draw [fill] (0,0) circle [radius=0.07];
\node [above] at (-1,1) {\scriptsize $i$};
\node [above] at (1,1) {\scriptsize $i'$};
\node [below] at (0,0) {$\Alg$};
\node [right] at (2,0.5) {$\ds \;=\; Y_{i}^{i'}[\Alg]\;,$};
\end{tikzpicture}
\end{equation}
As the result, equation (\ref{KS3}) become
\begin{equation}\label{KS4}
K_{12;\aa\bb}
\underbrace{ L_{\alpha,\beta}^{(2)} Y_\alpha^{(\bb)} L_{\beta,\alpha}^{(1)} Y_\beta^{(\aa)}}_{\tilde{s}_L}\;=\;
\underbrace{Y_\beta^{(\aa)} L_{\alpha,\beta}^{(1)} Y_\alpha^{(\bb)} L_{\beta,\alpha}^{(2)}}_{\tilde{s}_R} K_{12;\aa,\bb}
\end{equation}
Graphical representation for re-defined $\tilde{s}_L$ and $\tilde{s}_R$ is
\begin{equation}
\begin{tikzpicture}
\draw [thick, blue] (0,0) -- (6,0);
\draw [-stealth, thick] (6,2) -- (2,0) -- (0,1);
\draw [-stealth, thick] (5,2) -- (4,0) -- (3,2);
\draw [fill] (3.6,0.8) circle [radius=0.05];
\draw [fill] (4.667,1.333) circle [radius=0.05];
\draw [fill] (2,0) circle [radius=0.05];
\draw [fill] (4,0) circle [radius=0.05];
\node [above] at (5,2) {\scriptsize $\alpha$};
\node [above] at (6.2,2) {\scriptsize $\beta$};
\node [below] at (4,0) {\scriptsize $\Alg_q^{(\bb)}$};
\node [below] at (2,0) {\scriptsize $\Alg_q^{(\aa)}$};
\node [right] at (4.7,1.2) {\scriptsize $\Alg_{q^2}^{(2)}$};
\node [left] at (3.5,0.85) {\scriptsize $\Alg_{q^2}^{(1)}$};
\node [left] at (-1,1) {$\tilde{s}_L\;=\;$};
\end{tikzpicture}
\end{equation}
and
\begin{equation}
\begin{tikzpicture}
\draw [thick, blue] (0,0) -- (6,0);
\draw [-stealth, thick] (6,1) -- (4,0) -- (0,2);
\draw [-stealth, thick] (3,2) -- (2,0) -- (1,2);
\draw [fill] (1.333,1.333) circle [radius=0.05];
\draw [fill] (2.4,0.8) circle [radius=0.05];
\draw [fill] (2,0) circle [radius=0.05];
\draw [fill] (4,0) circle [radius=0.05];
\node [above] at (3,2) {\scriptsize $\alpha$};
\node [above] at (6.2,1) {\scriptsize $\beta$};
\node [below] at (4,0) {\scriptsize $\Alg_{q}^{(\aa)}$};
\node [below] at (2,0) {\scriptsize $\Alg_{q}^{(\bb)}$};
\node [right] at (2.4,0.9) {\scriptsize $\Alg_{q^2}^{(1)}$};
\node [left] at (1.333,1.2) {\scriptsize $\Alg_{q^2}^{(2)}$};
\node [left] at (-1,1) {$\tilde{s}_R\;=\;$};
\end{tikzpicture}
\end{equation}
This picture evidently defines $K$-matrix for the $C$-type reflection equation with "short" algebras $\Alg_{q}$ on the boundary and with "long" algebras $\Alg_{q^2}$ in the bulk. 

\subsection{Reflection equations}

Two $3d$ reflection equations in our space indices notations are 
\begin{equation}\label{Refl-B}
\begin{array}{l}
\ds K_{23;ef}^{}K_{13;cd}^{}R_{cbf}^{-1}R_{fda}^{}K_{12;ab}^{}R_{ace}^{-1}R_{ebd}^{}\;=\\
\\
\ds =\; R_{dbe}^{-1}R_{eca}^{} K_{12;ab}^{} R_{adf}^{-1}R_{fbc}^{} K_{12;cd}^{} K_{23;ef}^{}
\end{array}
\end{equation}
and
\begin{equation}\label{Refl-C}
\begin{array}{l}
\ds K_{12;ab}^{} K_{45;ac}^{} R_{531}^{}R_{146}^{-1} K_{36;bc}^{} R_{652}^{}R_{234}^{-1}\;=\\
\\
\ds =\; R_{432}^{}R_{256}^{-1} K_{36;bc}^{} R_{641}^{}R_{135}^{-1} K_{45;ac}^{} K_{12;ab}^{}\;.
\end{array}
\end{equation}
These equations can be obtained as the associativity conditions for relations (\ref{KS3}) and (\ref{KS4}) respectively. Comparing the structure of the indices, the reader can verify that equation (\ref{Refl-2}) can be seen as a particular case of equation (\ref{Refl-B}).

\subsection{Torus with the mirror}

Turn now to the integrable system on the torus with $M=N$:
\begin{equation}\label{Mir}
\begin{tikzpicture}
\draw [-stealth, thick] (0,-0.5) -- (0,2.5); 
\draw [-stealth, thick] (1,-0.5) -- (1,2.5);
\draw [-stealth, thick] (2,-0.5) -- (2,2.5);
	\node [below] at (0,-0.5) {\scriptsize $\alpha_{1}$};
	\node [below] at (1,-0.5) {\scriptsize $\alpha_{2}$};
	\node [below] at (2,-0.5) {\scriptsize $\alpha_{3}$};
\draw [-stealth, thick] (-0.5,0) -- (2.5,0);
\draw [-stealth, thick] (-0.5,1) -- (2.5,1);
\draw [-stealth, thick] (-0.5,2) -- (2.5,2);
	\node [left] at (-0.5,0) {\scriptsize $\beta_{1}$};
	\node [left] at (-0.5,1) {\scriptsize $\beta_{2}$};
	\node [left] at (-0.5,2) {\scriptsize $\beta_{3}$};
\draw [fill] (0,0) circle [radius=0.05];\draw [fill] (0,1) circle [radius=0.05];\draw [fill] (0,2) circle [radius=0.05];
\draw [fill] (1,0) circle [radius=0.05];\draw [fill] (1,1) circle [radius=0.05];\draw [fill] (1,2) circle [radius=0.05];
\draw [fill] (2,0) circle [radius=0.05];\draw [fill] (2,1) circle [radius=0.05];\draw [fill] (2,2) circle [radius=0.05];
\node [right] at (4,1) {$\to$};
\end{tikzpicture}\qquad
\begin{tikzpicture}
\draw [-stealth, thick] (0,-0.5) -- (0,2.5); 
\draw [-stealth, thick] (1,-0.5) -- (1,2.5);
\draw [-stealth, thick] (2,-0.5) -- (2,2.5);
	\node [below] at (0,-0.5) {\scriptsize $\alpha_{1}$};
	\node [below] at (1,-0.5) {\scriptsize $\alpha_{2}$};
	\node [below] at (2,-0.5) {\scriptsize $\alpha_{3}$};
\draw [-stealth, thick] (-0.5,0) -- (2.5,0);
\draw [-stealth, thick] (-0.5,1) -- (2.5,1);
\draw [-stealth, thick] (-0.5,2) -- (2.5,2);
	\node [left] at (-0.5,0) {\scriptsize $\beta_{1}$};
	\node [left] at (-0.5,1) {\scriptsize $\beta_{2}$};
	\node [left] at (-0.5,2) {\scriptsize $\beta_{3}$};
\draw [fill] (0,0) circle [radius=0.05];\draw [fill] (0,1) circle [radius=0.05];\draw [fill] (0,2) circle [radius=0.05];
\draw [fill] (1,0) circle [radius=0.05];\draw [fill] (1,1) circle [radius=0.05];\draw [fill] (1,2) circle [radius=0.05];
\draw [fill] (2,0) circle [radius=0.05];\draw [fill] (2,1) circle [radius=0.05];\draw [fill] (2,2) circle [radius=0.05];
\fill [blue!4!, path fading = east] (-1,-1) -- (3,3) -- (3,-1) -- cycle;
\fill [blue!4!, path fading = south] (-1,-1) -- (3,3) -- (3,-1) -- cycle;
\draw [blue, thin] (-1,-1) -- (3,3);
\end{tikzpicture}
\end{equation}
The diagonal line here stays as the mirror, the algebras on north-west corner are identified with the algebras on the south-east one. In Korepanov's formulation $X_{\alpha_i,\beta_j}=X_{\alpha_i,\beta_j}[\Alg_{q}^{(i,j)}]$, the symmetry is $\Alg_q^{(i,j)}\equiv\Alg_q^{(j,i)}$, $i\neq j$, and $X_{\alpha_i,\beta_i}=X_{\alpha_i,\beta_i}[\Alg_{q^2}^{(i)}]$ on the boundary.

In classics, the spectral determinants $J(\lambda,\mu)$ with the mirror symmetry and with the modified $\Alg_q$ in the bulk and $\Alg_{q^2}$ on the boundary brackets still are in involution. This can be verified in e.g. Maple or Mathematica for relatively small $N$.

The classical limit allows one to transform the spectral determinant (\ref{det}) to the transfer matrix (\ref{Tmat-s}) since the algebras are Abelian. The resulting transfer matrix (\ref{Tmat-s}), defined on the torus with the mirror symmetry, literally becomes at the same time the transfer matrix (\ref{Tmat}).
Geometrically, the construction of the transfer matrix (\ref{Tmat}) is the picture (\ref{Mir}) bended along the mirror line and folded into the half-plane. 

A natural question arisen it the quantum case is: which half-plane must stand above and which half-plane must stand below. The answer to this question is Statement \ref{prop1}: it does not matter, both orderings provide the same result. Thus, the alternative view to the system described at the beginning of this paper is the following.
\begin{itemize}
\item One can start with the transfer-matrix $T(\lambda,\mu)$ defined on the torus, eqs. (\ref{Tmat-s0},\ref{Tmat-s}), with the local algebras $\Alg_{q}^{(i,j)}$ with $i\neq j$, and with $\Alg_{q^2}^{(i)}$ on diagonal.
\item In this matrix, all elements of $\Alg_{q}^{(i,j)}$ with $i<j$ are to be placed to the left of all elements of $\Alg_{q}^{(i,j)}$ with $i>j$.
\item After this, one imposes the identification $\Alg_{q}^{(j,i)}=\Alg_{q}^{(i,j)}$, $i<j$.
\item The result is the transfer-matrix $T(\lambda,\mu)$ with the boundary, given by eq. (\ref{Tmat}) or (\ref{Tmat1}).
\end{itemize}

\section{Summary}

We present in this paper a completely integrable quantum-mechanical lattice system defined on a half-plane with a boundary.
The integrability is provided by auxiliary tetrahedra equations and by a specific version of an auxiliary tetrahedral reflection equation.

The system described has no straightforward quantum group classification.

Our system can be interpreted as an integrable system on a torus with a pairwise identification of different local components of the algebra of observables. There was a straightforward idea to look for a projection operator $\mathcal{P}\:\Alg_{q^2}\otimes\Alg_{q^2}\to\Alg_q$ such that
\begin{equation}
\mathcal{P} \; T_{\textrm{torus}}(\lambda,\mu) \;=\;
T_{\textrm{half-plane}}(\lambda,\mu) \; \mathcal{P}\;,
\end{equation}
but such attempt seems to be failed.

We guess, the approach developed in this paper can be extended to more general boundaries of $B$ type (situation with $C$-type remains unclear), however it requires additional algebraical exercises. Also, an open problem is spectral equations (an analogue of the nested Bethe Ansatz equations for the models on the torus) for the system described.

\noindent
\textbf{Acknowledgements.} The author would like to thank R. Kashaev, V. Bazhanov, V. Mangazeev and A. Kirillov for valuable discussions.


\begin{thebibliography}{99}

\bibitem{Cher:84}
I. V. Cherednik, \emph{``Factorizing particles on a half-line and root systems''}, Theor Math Phys \textbf{61} (1984) 977–983

\bibitem{Skl:88}
E. K. Sklyanin, \emph{``Boundary conditions for integrable quantum systems''},
J. Phys. A: Math. Gen. \textbf{21} (1988) 2375

\bibitem{IK:97}
A. P. Isaev and P. P. Kulish,
\emph{``Tetrahedron Reflection Equations''},
Modern Physics Letters A \textbf{12} No. 06 (1997) 427-437




\bibitem{KO:2012}
A. Kuniba ana M. Okado, 
\emph{``Tetrahedron and $3D$ reflection equations from quantized algebra of functions''},
J. Phys. A: Math. Theor. \textbf{45} (2012) 465206


\bibitem{KOY:2019}
A. Kuniba, M. Okado and A. Yoneyama,
\emph{``Reflection $\boldsymbol{K}$ matrices associated with an Onsager coideal of 
$U_p(A^{(1)}_{n-1})$, $U_p(B^{(1)}_n)$,$U_p(D^{(1)}_n)$ and $U_p(D^{(2)}_{n+1})$''},
J. Phys. A: Math. Theor. \textbf{52} (2019) 375202

\bibitem{IKT:2024}
R. Inoue, A. Kuniba amd Y. Terashima,
\emph{``Quantum Cluster Algebras and 3D Integrability: Tetrahedron and 3D Reflection Equations''},
International Mathematics Research Notices \textbf{16} (2024) 11549-11581


\bibitem{Zam:80}
A. B. Zamolodchikov, 
\emph{``Tetrahedra equations and integrable systems in three-dimensional space''},
Journal of Experimental and Theoretical Physics \textbf{52} (1980) 325



\bibitem{Maillard:1982}
M. T. Jaekel and J. M. Maillard,
\emph{``Symmetry relations in exactly soluble models''}, J. Phys. A: Math. Gen. 15 (1882) 1309

\bibitem{BS:1984}
V. V.  Bazhanov and Yu. G. Stroganov. 
\emph{“Free fermions on a three-dimensional lattice and tetrahedron equations.”}
Nuclear Physics \textbf{B230} (1984) 435-454

\bibitem{BS:2006}
V. V. Bazhanov and S. M. Sergeev, \emph{``Zamolodchikov's tetrahedron equation and hidden structure of quantum groups''}, J. Phys. A: Math. Gen. 39 (2006) 3295-3310

\bibitem{K:95}
I. G. Korepanov, \emph{Algebraic integrable systems, 2+1 dimensional models on wholly discrete space-time, and inhomogeneous models on 2-dimensional statistical physics}, Adv. PhD. Thesis, arXiv:solv-int/9506003, 1995

\bibitem{S:2006}
S. Sergeev, \emph{Quantum curve in q-oscillator model}, 
International Journal of Mathematics and Mathematical Sciences
Volume 2006, Article ID 92064, Pages 1–31

\bibitem{S:2009}
S. Sergeev, \emph{Super-tetrahedra and super-algebras}, J. Math. Phys. 50 (2009)
083519. 

\bibitem{KV:91}
M. M. Karpanov and V. A. Voevodsky, \emph{``$2$-categories and Zamolodchikov tetrahedra equations''}, 
Algebraic groups and their generalisations: quantum and infinite-dimensional methods (University Park,
PA, 1991), Proc. Sympos. Pure Math., vol. 56, Amer. Math. Soc., Providence, RI, 1994, pp. 177-259.

\bibitem{KS:93}
D. Kazhdan and Y. Soibelman, \emph{``Representations of the quantized function algebras, $2$-categories and Zamolodchikov tetrahedra equation''}, The Gel'fand Mathematical Seminars, 1990-1992, Birkhauser, Boston, MA, 1993, pp. 163-171

\bibitem{Soi:90}
Ya. S. Soibelman, \emph{``Algebra of functions on a compact quantum group and its representations''},
Algebra i Analiz \textbf{2} (1990) no. 1, pp. 190-212



\end{thebibliography}
\end{document}